\documentclass{book}
\usepackage{makeidx}
\usepackage{setspace}%,subfigure
\usepackage{Generic}
\usepackage{amssymb}
\usepackage{graphicx,epsfig}
\usepackage{amsmath}
\usepackage{textcomp}
\usepackage{lscape}
\usepackage{amsthm}
\usepackage{bm}
\usepackage{colordvi}

\newcommand{\bgreek}[1]{\mbox{\boldmath$#1$\unboldmath}}

\begin{document}

\pagenumbering{arabic}

\title{Spin relaxation and spin dynamics in semiconductors}

\author{J. Fabian and M. W. Wu}

%
%\pacs{72.25.Rb, 72.25.Dc, 76.30.Pk}

\maketitle

The spin of conduction electrons decays due to the combined effect of spin-orbit coupling
and momentum scattering. The spin-orbit coupling couples the spin to the electron momentum,
that is randomized by momentum scattering off of impurities and phonons. Seen from the perspective
of the electron spin, the spin-orbit coupling gives a spin precession, while momentum
scattering makes this precession randomly fluctuating, both in magnitude and orientation.

The specific mechanisms for the spin relaxation of conduction electrons were proposed by
Elliott \cite{Elliott1954:PR} and Yafet \cite{Yafet:1963},
for conductors with a center of inversion symmetry, and by
D'yakonov and Perel' \cite{Dyakonov1972:SPSS}, for conductors without an inversion center. In p-doped semiconductors
there is in play another spin relaxation mechanism, due to Bir, Aronov, and Pikus \cite{Bir1976:SPJETP}. As this
has a rather limited validity we do not describe it here. More details can be found in reviews
\cite{Meier:1984, Fabian1999:JVST, Zutic2004:RMP, Fabian2007:APS, wu-review}.

Before we discuss the two main mechanisms, we introduce a toy model that captures the
 relevant physics of spin relaxation without resorting explicitly to quantum mechanics:
 \emph{the electron spin in a randomly fluctuating magnetic field}. We will find
 certain universal qualitative features of the spin relaxation and dephasing in physically
important situations.

The next part of this review covers the experimental as well as computational status
of the field, discussing the spin relaxation in semiconductors under varying conditions
such as temperature and doping density.

\section{Toy model: the electron spin in a fluctuating magnetic field. \label{sec:toy}}

\begin{figure}
\centerline{\psfig{file=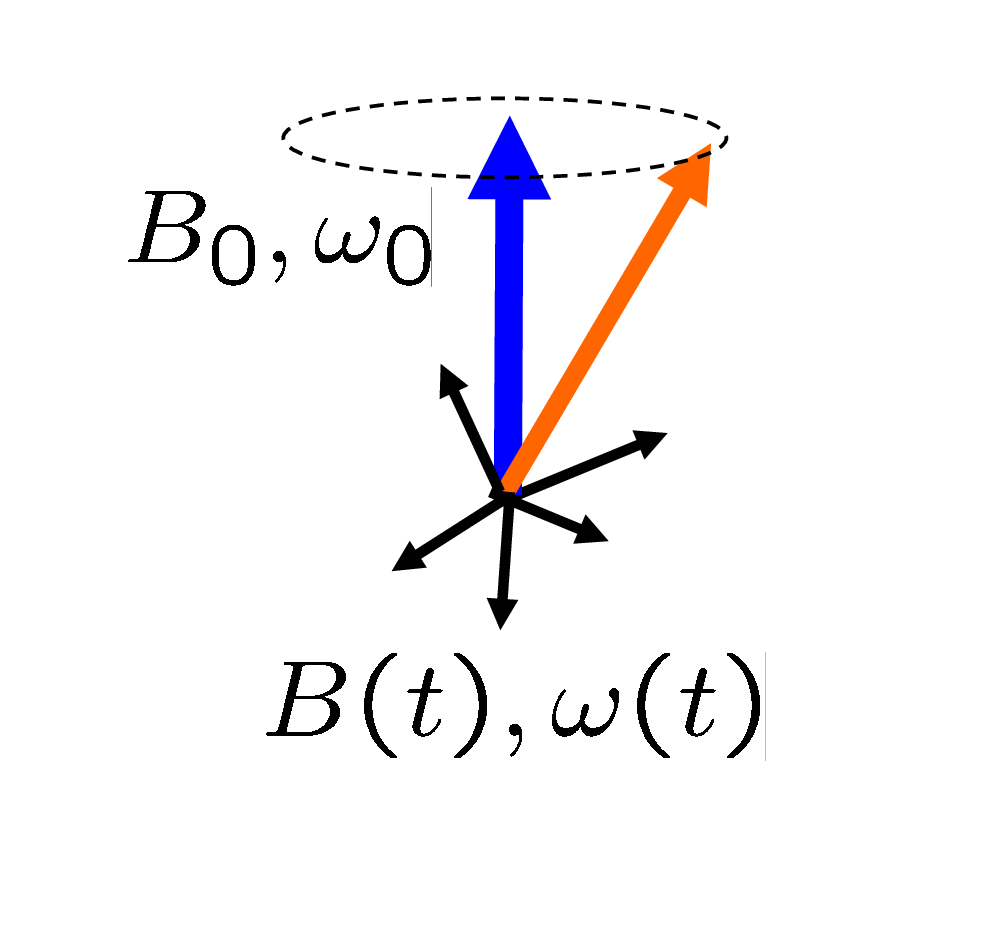,width=0.3\linewidth}}
\caption{Electron spin precesses about the static $B_0$ field along
$\bf z$. The randomly fluctuating magnetic field ${\bf B}(t)$ causes
spin relaxation and spin dephasing.}
\label{fig:fluctuating_spin}
\end{figure}

Consider an electron spin $\bf S$ (or the corresponding magnetic moment) in the presence of an external time-independent magnetic field ${\bf B}_0 = B_0{\bf z}$
giving rise to the Larmor precession frequency $\bgreek{\omega}_0 = \omega_0 {\bf z} $,
and a fluctuating time-dependent field $ {\bf B}(t)$ giving the Larmor frequency ${\bgreek{\omega}}(t)$;
see Fig. \ref{fig:fluctuating_spin}.
We assume that the field fluctuates about zero and is correlated on the time scale of $\tau_c$:
\begin{equation} \label{eq:fluctuating-field11}
\overline{\omega (t)} = 0, \quad \overline{\omega_\alpha(t) \omega_{\beta}(t')} =\delta_{\alpha
\beta}\overline{\omega_\alpha^2}e^{-|t-t'|/\tau_c}.
\end{equation}
Here $\alpha$ and $\beta$ denote the cartesian coordinates and the overline
denotes averaging over different random realizations ${\bf B}(t)$. We will see later that
such fluctuating fields arise quite naturally in the context of the electron spins in solids.

The following description applies equally to the classical magnetic moment
described by the vector $\bf S$ as well as to the quantum mechanical spin
whose expectation value is $\bf S$. Writing out the torque equation, $\dot{\bf S} =
\bgreek{\omega} \times {\bf S}$, we get the following equations of motion:
\begin{eqnarray}
\dot{S}_x & = & - \omega_0 S_y + \omega_y(t) S_z - \omega_{z}(t) S_y, \\
\dot{S}_y & = & \omega_0 S_x - \omega_x(t) S_z + \omega_z(t) S_x, \\
\dot{S}_z & = & \omega_x(t) S_y - \omega_y(t) S_x.
\end{eqnarray}
These equations are valid for one specific realization of $\bgreek{\omega}(t)$.
Our goal is to find instead effective equations for the time evolution of the
average spin, ${\overline{\bf S}}(t)$, given the ensemble of Larmor
frequencies $\bgreek{\omega}(t)$.

It is convenient to introduce the complex ``rotating'' spins $S_{\pm}$ and Larmor frequencies
$\omega_{\pm}$ in the $(x,y)$ plane:
\begin{eqnarray}
S_+ & =& S_x + i S_y, \quad S_-  =  S_x - i S_y, \\
\omega_+ & =& \omega_x + i \omega_y, \quad \omega_-  =  \omega_x - i \omega_y.
\end{eqnarray}
The inverse relations are
\begin{eqnarray}
S_x & =& \frac{1}{2}\left ( S_+ + S_-\right ), \quad S_y  = \frac{1}{2i} \left (S_+ - S_-\right), \\
\omega_x & =& \frac{1}{2}\left ( \omega_+ + \omega_- \right ) , \quad \omega_y  =
\frac{1}{2i} \left ( \omega_+ -  \omega_- \right).
\end{eqnarray}
The equations of motion for the spin set $(S_+, S_-, S_z)$ are,
\begin{eqnarray}
\dot{S}_+ & = & i \omega_0 S_+ + i \omega_z S_+ - i \omega_+ S_z, \\
\dot{S}_- & = & - i \omega_0 S_- - i\omega_z S_- + i \omega_- S_z, \\
\dot{S}_z & = & - (1/2i) \left ( \omega_+ S_- - \omega_- S_+ \right ).
\end{eqnarray}
In the absence of the fluctuating fields the spin $S_+$ rotates in the
complex plane anticlockwise (for $\omega_0 > 0$), while $S_-$ clockwise.

The precession about $B_0$ can be factored out
by applying the ansatz, \footnote{This is analogous to going to the interaction
picture when dealing with a quantum mechanical problem of that type.}:
\begin{equation}
S_{\pm} = s_{\pm}(t) e^{\pm i \omega_0 t}.
\end{equation}
Indeed, it is straightforward to find the time evolution of the set $(s_+, s_-, s_z\equiv S_z)$:
\begin{eqnarray} \label{eq:rotating-frame21}
\dot{s}_{+} & = &  i \omega_z s_{+} - i \omega_+ s_z e^{-i \omega_0 t}, \\
\label{eq:rotating-frame22}
\dot{s}_{-} & = & - i\omega_z s_{-} + i \omega_- s_z e^{ i \omega_0 t} , \\
\label{eq:rotating-frame23}
\dot{s}_{z} & = & - (1/2i) \left ( \omega_+ s_{-} e^{-i \omega_0 t} - \omega_- s_{+} e^{i \omega_0 t} \right ).
\end{eqnarray}
The penalty for transforming into this ``rotating frame'' is the
appearance of the phase factors $\exp(\pm i \omega_0 t)$.

The solutions of Eqs. \ref{eq:rotating-frame21}, \ref{eq:rotating-frame22}, and
  \ref{eq:rotating-frame23}, can be written in terms of the integral equations, 
\begin{eqnarray}
{s}_{+}(t) & = & s_{+}(0) + i \int_0^t dt' \omega_z(t') s_{+}(t')
 - i \int_0^t dt' \omega_+(t')  s_z (t')  e^{-i \omega_0 t'}, \\
 {s}_{-}(t) & = & s_{-}(0) - i \int_0^t dt' \omega_z(t') s_{-}(t')
 + i \int_0^t dt' \omega_-(t')  s_z (t')  e^{i \omega_0 t'}, \\
 {s}_{z}(t) & = & s_{z}(0) -\frac{1}{2i} \int_0^t dt' \left [ \omega_+(t') s_{-}(t')
 e^{-i \omega_0 t'} -  \omega_-(t') s_{+}(t')
 e^{i \omega_0 t'} \right ].
 \end{eqnarray}
We should now substitute the above solutions back into Eqs. \ref{eq:rotating-frame21}, \ref{eq:rotating-frame22}, and  \ref{eq:rotating-frame23}. The corresponding expressions become
rather lengthy, so we demonstrate the procedure on the $s_+$ component only. We get,
\begin{eqnarray} \nonumber
\dot{s}_+(t) =&& i \omega_z(t) s_{+}(0) - \omega_z(t)
 \int_0^t dt' \omega_z(t') s_{+}(t') +
 \omega_z(t) \int_0^t dt' \omega_+(t')  s_z (t')  e^{-i \omega_0 t'} \\
&&-i \omega_+(t) e^{-i\omega_0 t} s_{z}(0)
+ \frac{1}{2}e^{-i\omega_0 t}  \omega_+(t) \int_0^t dt' \left [ \omega_+(t') s_{-}(t')
 e^{-i \omega_0 t'} -  \omega_-(t') s_{+}(t')
 e^{i \omega_0 t'} \right ]. \label{eq:splus21}
\end{eqnarray}
The reader is encouraged to write the analogous equations for $\dot{s}_z$ (that for $\dot{s}_-$
is easy to write since $s_- = s_+^*$).

We now make two approximations. First, we assume that the fluctuating field is rather weak and
stay in {\it the second order} in $\omega$.\footnote{More precisely, we assume
that $|\omega(t)| \tau_c \ll 1$, so that the spin does not fully precess about the fluctuating
field before the field makes a random change.} This allows us to factorize the
averaging over the statistical realizations of the field,
\begin{equation} \label{eq:Born}
\overline{ \omega(t) \omega(t') s(t')} \approx \overline{\omega(t) \omega(t')}\,\, \overline{s(t')}
\end{equation}
as the spin changes only weakly over the time scale, $\tau_c$, of the changes of the
fluctuating fields. This approximation is called the \emph{Born approximation}, alluding to the
analogy with the second-order time-dependent perturbation theory in quantum mechanics.
Going beyond the Born approximation one would need to execute complicated
averaging schemes of the product in Eq. \ref{eq:Born}, since $s(t)$ in general
depends on $\omega(t' \le t)$.

As the second assumption, we consider a ``coarse-grained" time evolution, meaning that
we are effectively averaging $s(t)$ over the time scale of the correlation time
$\tau_c$; we are interested in times $t$ much greater than $\tau_c$. That allows
us to approximate,
\begin{equation} \label{eq:Markov}
\int_0^{t \gg \tau_c} dt' \,\overline{\omega(t) \omega(t')}\,\, \overline{s(t')} \approx \int_0^{t\gg \tau_c}
dt'\, \overline{\omega(t) \omega(t')}\,\, \overline{s(t)},
\end{equation}
since the correlation function  $\overline{\omega(t) \omega(t')}$ is significant in the
time interval of $|t-t'| \approx \tau_c$ only. The above approximation makes clear
that the spin $s(t)$ is the representative coarse-grained (running-averaged) spin of the time interval
$(t-\tau_c, t)$. Equation \ref{eq:Markov} is a realization of the {\it Markov approximation}.
The physical meaning is that the spin $s$ varies only slowly on the time scale of $\tau_c$ over which the
correlation of the fluctuating fields is significant. We then need to restrict
ourselves to the time scales $t$ larger than the correlation time $\tau_c$.
In effect, we will see that in this approximation
\emph{the rate of change of the spin at a given time depends on the spin at that time, not on the
previous history of the spin. }

Applying the Born-Markov approximation to Eq.  \ref{eq:splus21} we obtain for the average
spin $\overline{s}_+$ the following
time evolution equation:\footnote{The initial values of the spin, $s(0)$, are fixed and not affected by
averaging.}
\begin{eqnarray} \nonumber
\dot{\overline{s}}_+ =&& i \overline{\omega_z(t)} s_{+}(0) -
 \int_0^t dt' \overline{\omega_z(t) \omega_z(t')}\,\, \overline{s_{+}(t)} +
  \int_0^t dt' \overline{\omega_z(t) \omega_+(t')}  e^{-i \omega_0 t'}\overline{s_z (t)} \\
&&-i \overline{\omega_+(t)} e^{-i\omega_0 t} s_{z}(0)
+ \frac{1}{2}e^{-i\omega_0 t}  \int_0^t dt' \left [ \overline{\omega_+(t) \omega_+(t')}
e^{-i \omega_0 t'} \overline{s_{-}(t)}
  -  \overline{\omega_+(t) \omega_-(t')} e^{i \omega_0 t'} \overline{s_{+}(t)}  \right ].
\end{eqnarray}
Using the rules of Eqs. \ref{eq:fluctuating-field11} the above simplifies to
\begin{equation}
{\dot{\overline{s}}_+} =- \overline{\omega_z^2} \int_0^t dt' e^{-(t-t')/\tau_c} \overline{s_{+}(t)} +
 \frac{1}{2}e^{-i\omega_0 t}  \int_0^t dt' \left [ (\overline{\omega_x^2} - \overline{\omega_y^2})
e^{-i \omega_0 t'} \overline{s_{-}(t)}
  - (\overline{\omega_x^2} + \overline{\omega_y^2})  e^{i \omega_0 t'} \overline{s_{+}(t)}
  \right ]e^{-(t-t')/\tau_c} .
\end{equation}
Since we consider the times $t \gg \tau_c$, we can approximate
\begin{equation}
\int_0^t dt' e^{-(t-t')/\tau_c} \approx \int_{-\infty}^t dt' e^{-(t-t')/\tau_c} = \tau_c.
\end{equation}
Similarly,
\begin{equation}
\int_0^t dt' e^{-(t-t')/\tau_c} e^{-i\omega_0 (t \pm t')} \approx
\int_{-\infty}^t dt' e^{-(t-t')/\tau_c} e^{-i\omega_0 (t \pm t')} =
\tau_c \frac{1 \mp i \omega_0 \tau_c}{1 + \omega_0^2 \tau_c^2}.
\end{equation}
The imaginary parts induce the precession of $s_{\pm}$, which is equivalent
to shifting (renormalizing) the Larmor frequency $\omega_0$. The relative
change of the frequency is $(\omega \tau_c)^2$ which is assumed much smaller
than one by our Born approximation. We thus keep the real parts only
and obtain,
\begin{equation}
\dot{\overline{s}}_+ = -\overline{\omega_z^2} \tau_c \overline{s_+} +
\frac{1}{2}\frac{\tau_c}{1+ \omega_0^2 \tau_c^2}
\left [(\overline{\omega_x^2} - \overline{\omega_y^2})\overline{s_-}e^{-2i \omega_0 t} -
(\overline{\omega_x^2} +\overline{\omega_y^2}) \overline{s_+} \right ].
\end{equation}
Using the same procedure (or simply using $s_- = s_+^*$) we would arrive for the analogous
equation for $s_-$:
\begin{equation}
\dot{\overline{s}}_- = -\overline{\omega_z^2} \tau_c \overline{s_-} +
\frac{1}{2}\frac{\tau_c}{1+ \omega_0^2 \tau_c^2}
\left [(\overline{\omega_x^2} - \overline{\omega_y^2})\overline{s_+}e^{2i\omega_0 t} -
(\overline{\omega_x^2} + \overline{\omega_y^2}) \overline{s_-} \right ].
\end{equation}
Similarly,
\begin{equation}
\dot{\overline{s}}_z = -(\overline{\omega_x^2} + \overline{\omega_y^2}) 
\frac{\tau_c}{1 + \omega_0^2 \tau_c^2} \overline{s_z}
\end{equation}

\emph{For the rest of the section we omit the overline on the symbols for the spins, so
that $S$ will mean the average spin.} Returning back to our rest frame of the spins rotating
with frequency $\omega_0$, we get
\begin{eqnarray}
\dot{S}_+ & = & i \omega_0 S_+ - \overline{\omega_z^2} \tau_c
\frac{1}{2}\frac{\tau_c}{1+ \omega_0^2 \tau_c^2}
\left [(\overline{\omega_x^2} - \overline{\omega_y^2})S_-  -
(\overline{\omega_x^2} + \overline{\omega_y^2}) S_+ \right ], \\
\dot{S}_- & = & i \omega_0 S_- - \overline{\omega_z^2} -
\frac{1}{2}\frac{\tau_c}{1+ \omega_0^2 \tau_c^2}
\left [(\overline{\omega_x^2} - \overline{\omega_y^2})S_+  -
(\overline{\omega_x^2} + \overline{\omega_y^2}) \overline{S_-} \right ], \\
\dot{S}_z & = & -(\overline{\omega_x^2} + \overline{\omega_y^2}) 
\frac{\tau_c}{1 + \omega_0^2 \tau_c^2} \overline{S_z}.
\end{eqnarray}
Finally, going back to $S_x$ and $S_y$:
\begin{eqnarray}
\dot{S}_x & = & - \omega_0 S_y - \overline{\omega_z^2} \tau_c {S_x}
- \frac{\tau_c}{1+ \omega_0^2 \tau_c^2}
\overline{\omega_y^2}\overline{S_x}\\
\dot{S}_y & = &  \omega_0 S_x - \overline{\omega_z^2} \tau_c {S_y}
- \frac{\tau_c}{1+ \omega_0^2 \tau_c^2}
(\overline{\omega_x^2}) {S_y}, \\
\dot{S}_z & = & -(\overline{\omega_x^2} + \overline{\omega_y^2})
 \frac{\tau_c}{1 + \omega_0^2 \tau_c^2} {S_z}.
\end{eqnarray}
We can give the above equation a more conventional form, by introducing two types
of the spin decay times. First, we define the {\it spin relaxation time} $T_1$ by,
\begin{equation}
\frac{1}{T_1} = \left (\overline{\omega_x^2} + 
\overline{\omega_y^2}\right ) \frac{\tau_c}{1+\omega_0^2 \tau_c^2},
\end{equation}
and the spin dephasing times $T_2$ by
\begin{eqnarray}
\frac{1}{T_{2x}} & = & \overline{\omega_z^2} \tau_c + \frac{\overline{\omega_y^2}\tau_c}{1+\omega_0^2 \tau_c^2}, \\
\frac{1}{T_{2y}} & = & \overline{\omega_z^2} \tau_c + \frac{\overline{\omega_x^2}\tau_c}{1+\omega_0^2 \tau_c^2}.
\end{eqnarray}
We then write:
\begin{eqnarray}
\dot{S}_x & =&  - \omega_0 S_y - \frac{S_x}{T_{2x}}, \\
\dot{S}_y & = &  \omega_0 S_x - \frac{S_y}{T_{2y}}, \\
\dot{S}_z & = &  - \frac{S_z}{T_{1}}.
\end{eqnarray}
Our fluctuating field is effectively at infinite temperature, at which the average
value for the spin in a magnetic field is zero. A more general spin dynamics is
\begin{eqnarray}
\dot{S}_x & =&  - \omega_0 S_y - \frac{S_x}{T_{2x}}, \\
\dot{S}_y & = &  \omega_0 S_x - \frac{S_y}{T_{2y}}, \\
\dot{S}_z & = &  - \frac{S_z- S_{0z}}{T_{1}}.
\end{eqnarray}
where $S_{0z}$ is the equilibrium value of the spin in the presence of the
static magnetic field of the Larmor frequency $\bgreek{\omega}_0$ at the
temperature at which the environmental fields giving rise to $\bgreek{\omega}(t)$
are in equilibrium. The above equations are called the \emph{Bloch equations}.

The spin components $S_x$ and $S_y$, which are perpendicular to the applieed
static field ${\bf B}_0$, decay exponentially on the time scales of $T_{2x}$
and $T_{2y}$, respectively. These times are termed
spin {\it dephasing} times, as they describe the loss of the \emph{phase} of the spin components
perpendicular to the static field ${\bf B}_0$. They are also often called {\it transverse}
times, for that reason. The time $T_1$ is termed the spin {\it relaxation} time, as it describes the (thermal)
relaxation of the spin to the equilibrium. During the spin relaxation in a static magnetic field
the energy is exchanged with the environment. In the language of statistical physics, the relaxation process
establishes the Boltzmann probability distribution for the system. Similarly, dephasing establishes
the ``random phases" postulate that says that there is no
correlation (coherence) among the degenerate states, such as the two transverse
spin orientations; in thermal equilibrium such states add incoherently.

For the sake of discussion consider an isotropic system in which
\begin{equation}
\overline{\omega_x^2} = \overline{\omega_y^2} = \overline{\omega_z^2} = \overline{\omega^2}.
\end{equation}
If the static magnetic field is weak, $\omega_0 \tau_c \ll 1$, the three
times are equal:
\begin{equation} \label{eq:spin-relaxation41}
T_{1\cdot} = T_{2x} = T_{2y} = \frac{1}{\overline{\omega^2} \tau_c}.
\end{equation}
There is no difference between the spin relaxation and spin dephasing. It is
at first sight surprising that the spin relaxation time is inversely proportional
to the correlation time. The more random the external field appears, the less the
spin decays. We will explain this fact below by the phenomenon of motional
narrowing.

In the opposite limit of large Larmor frequency, $\omega_0 \tau_c
\gg 1$, the spin relaxation rate vanishes,
\begin{equation}
\frac{1}{T_1} \approx \frac{\overline{\omega^2}}{\omega_0^2} \frac{1}{\tau_c}
\to   0,
\end{equation}
while the spin dephasing time is given by what is called \emph{secular broadening},
\begin{equation}
\frac{1}{T_2} \approx  \overline{\omega_z^2} \tau_c.
\end{equation}
If secular broadening is absent, the leading term in the dephasing
time will be, as in the relaxation,
\begin{equation}
\frac{1}{T_2} \approx \frac{\overline{\omega^2}}{\omega_0^2}
\frac{1}{\tau_c}.
\end{equation}
In this limit the spin dephasing rate is proportional to the
correlation rate, not to the correlation time.

In the cases in which there is no clear distinction
between $T_1$ and $T_2$, we often use the symbol
\begin{equation}
\tau_s = T_1 = T_2,
\end{equation}
to describe the \emph{generic spin relaxation}.

\subsubsection{Motional narrowing}

The surprising fact that, at low magnetic fields, the spin relaxation rate is proportional to
the correlation time of the fluctuating field (as opposed to its inverse), is explained by \emph{motional narrowing}. Consider the spin transverse to an applied magnetic field and assume that the field
has a single magnitude, but can randomly  switch directions, between up to down, leading to
a random precession of the spin clock and anticlockwise.  In effect, the spin phase executes a random
walk. A single step takes the time $\tau_c$, the correlation time of the fluctuating field. After
$n$ steps, that is, after the time $t=n\tau$, the standard deviation of the phase will be $\delta \phi= (\omega
\tau_c)\sqrt{n}$, the well known result for a random walk. We call the spin dephasing time
the time it takes for $\delta \phi \approx 1$. This happens after the time $\tau_s = \tau_c/(\omega \tau_c)^2$,
or $\tau_s = 1/(\omega^2 \tau_c)$, which is the result we obtained earlier from the Born-Markov approximation,
Eq. \ref{eq:spin-relaxation41}.

\subsection{Reversible dephasing, spin ensemble, random walk in inhomogeneous fields.}

Our previous calculation was carried out for a single spin in a fluctuating magnetic field. After the
decay of the spin components, the information about the original spin is irreversibly
lost as we have no information on the actual history of the fluctuating field.
Such an irreversible loss of spin is often termed spin \emph{decoherence}. We will see that
spin dephasing can be reversible. We will also see that a simple exponential
decay, of the type $\exp(-t/\tau_s)$,  is not a rule.

\subsubsection{Reversible spin dephasing: spin ensemble in spatially random magnetic field}

\begin{figure}
\centerline{\psfig{file=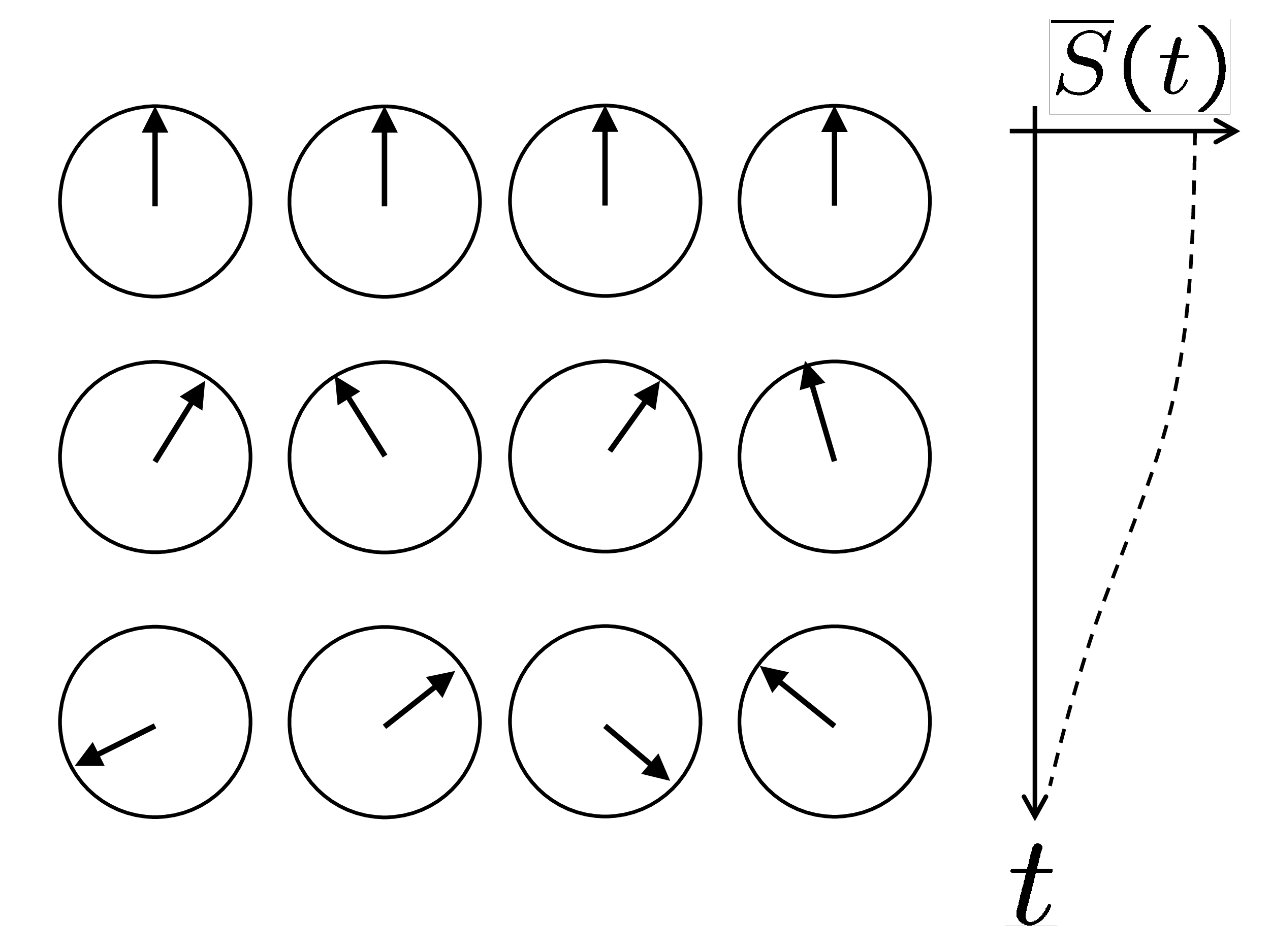,width=0.4\linewidth}}
\caption{Electron spin precesses about the static ${\bf B}_0$ field along
$\bf z$ perpendicular to this page.
The spatially fluctuating magnetic field $B_1{\bf z}$ causes
reversible spin spin dephasing.}
\label{fig:ensemble_spin}
\end{figure}

There are physically relevant cases in which the decay of spin is reversible. A typical
example is an {\it ensemble} of localized spins, each precessing about a static local magnetic field
${\bf B}_0 + {\bf B}_1$, with varying static components ${\bf B}_1$ (of zero average) giving rise to random
precession frequencies $\bgreek{\omega}_1$. See Fig. \ref{fig:ensemble_spin}.
Another important example is that of the conduction electrons
in noncentrosymmetric crystals, such as GaAs, in which the spin-orbit coupling acts
as a momentum dependent magnetic field. The spins of the electrons of different momenta precess
with different frequencies. We are interested in the total spin as the sum
of the individual spins.

Consider the external field along $z$ direction, and take the fluctuating frequencies
from the Gaussian distribution:
\begin{equation}
P(\omega_1) = \frac{1}{\sqrt{2\pi \delta\omega^2}} e^{-\omega_1^2/2\delta\omega^2},
\end{equation}
with zero mean and $\delta\omega_1^2$ variance. Denote the in-plane spin of
the electron $a$ \footnote{This discussions applies to nuclear spins
as well.} by $S_{x}^a$ and $S_{y}^a$. This spin precesses with the
frequency $\omega_0+ \omega_1^a$, leading to the time evolution for the rotating
spins
\begin{equation}
S_{\pm}^a(t) = S_{x}^a(t) \pm i S_{y}^a(t) = S_{\pm}^a(0) e^{\pm i \omega_0 t}  e^{\pm i \omega_1^a t}.
\end{equation}
Suppose at $t=0$ all the spins are lined up, that is, $S^a_{\pm}(0) = S_{\pm}(0)$.
The total spin $S_{\pm}(t)$ is the sum,
\begin{equation}
S_{\pm}(t) = \sum_s  S_{\pm}^a(t) = S_{\pm}(0) e^{\pm i \omega_0 t} \int_{-\infty}^{\infty} d\omega_1
P(\omega_1) e^{\pm i \omega_1 t}.
\end{equation}
Evaluating the Gaussian integral we get
\begin{equation}
S_{\pm}(t) = S_{\pm}(0) e^{\pm i \omega_0 t} e^{- \delta\omega_1^2 t^2/2}.
\end{equation}
The in-plane component vanishes after the time of about $1/\delta\omega_1$, but this
dephasing of the spin is reversible, since each individual spin preserves the memory
of the initial state. The disappearance of the spin is purely due to the statistical
averaging over an ensemble in which the individual spins have, after certain time,
random phases. This spin decay is not a simple exponential, but rather Gaussian.

\subsubsection{Spin echo}

\begin{figure}
\centerline{\psfig{file=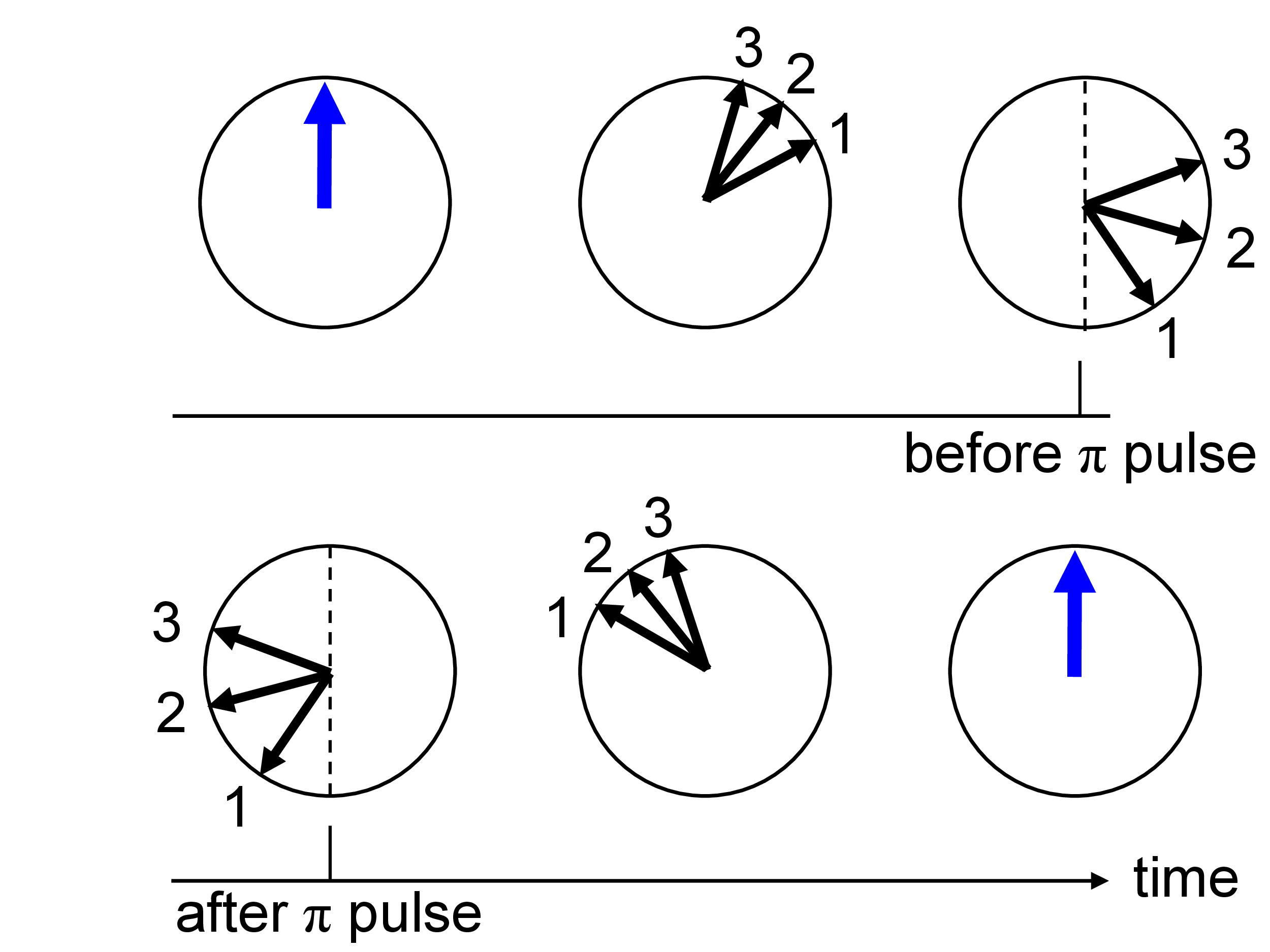,width=0.4\linewidth}}
\caption{Initially all the spins point up. Due to random precession frequencies the
spins soon point in different directions and the average spin vanishes. Applying
a $\pi$-pulse rotating the spins along the vertical axis makes the fastest spin the
slowest, and vice versa. After the fastest ones catch up again with the slowest, the
original value of the average spin is restored. Any reduction from the original value
is due to irreversible processes.
signal.}
\label{fig:spin_echo}
\end{figure}

We have seen that there are irreversible and reversible effects both present in spin dephasing. It turns
out that the reversible effects can be separated out by the phenomenon of the spin echo.
Figure \ref{fig:spin_echo} explains
this mechanism in detail. Suppose all the localized spins in our ensemble point in one
direction at time $t=0$. At a later
time, $t=T_\pi$, the spins dephase due to the inhomogeneities of the precession frequencies and
the total spin is small. We apply a short
pulse of an external magnetic field, the so called $\pi$ pulse, that rotates the spins along the
axis parallel to the original spin direction, mapping the spins as
$(S_x, S_y) \to (-S_x, S_y)$. At that moment the spins will still be dephased, but the one that is
the fastest is now the last, and the one that is the slowest, appears as the first. At the time
$t=2T_\pi$ all the spins catch on, producing a large spin signal along the original spin direction.
In reality this signal will be weaker than that at $t=0$ due to the presence of irreversible processes,
by $\exp(-2T_\pi/T_2)$. Important, reversible processes are not counted in $T_2$.

\subsubsection{Spin random walk in inhomogeneous magnetic field}

Another interesting situation appears when we consider the possibility that the
spin diffuses through a region of an inhomogeneous precession frequency.\footnote{The
inhomogeneity could be due to the magnetic field or to the g-factor.} We can model this
situation on the system of spins in one dimension, each spin jumping
in a time $\tau$ left or right. Suppose the precession frequencies vary in the
$x$ direction as
\begin{equation}
\omega(x) = \omega_0 + \omega' x.
\end{equation}
Here $\omega'$ is the gradient of $\omega$. As the electron diffuses, it precesses
with the precession frequency $\omega(t) = \omega[x(t)]$, given by the position of the
electron $x(t)$ at time $t$. The time evolution for an individual spin is
\begin{equation}
\dot{S}_+ = \omega(t) S_+(t) = \omega(t)S_{+}(0) + \omega(t) \int_0^\infty dt' \omega(t') S(t').
\end{equation}
We are interested in the averaged quantities over many realizations of the
random walk. We also consider the time scales much greater than the individual random
walks steps $\tau$.

The accumulated phase after $N$ steps or $t=N\tau$ time is the sum
\begin{equation}
\phi_N = x_1 + x_2 + ... + x_N = N \delta_1 + (N-1) \delta_1 + ... + \delta_N.
\end{equation}
Here $\delta_i = \pm 1$ is a random variable representing the random
step left or right. The variance of $\phi_N$ is
\begin{equation}
\sigma^2_{N} = \sum_i^{N-1} (N-i)^2 \approx \frac{1}{3}N^3.
\end{equation}
According to the central limit theorem, $\phi_N$ is distributed normally with the
above variance:
\begin{equation}
P(\phi_N) = \frac{1}{\sqrt{2\pi \sigma^2_N}} e^{-\phi_N^2/2\sigma^2_N}.
\end{equation}
We can now transform the dynamical equation into the ensemble averaging:
\begin{equation}
S_+(t) = S_+(0) e^{i\phi(t)} = \int_{-\infty}^{\infty} d\phi_N e^{i\phi(t)} P(\phi_N) = e^{-\omega_1^2 D t^3/3},
\end{equation}
where we denoted the diffusivity by $D = \tau/2$, for the unit length step.

\begin{figure}
\centerline{\psfig{file=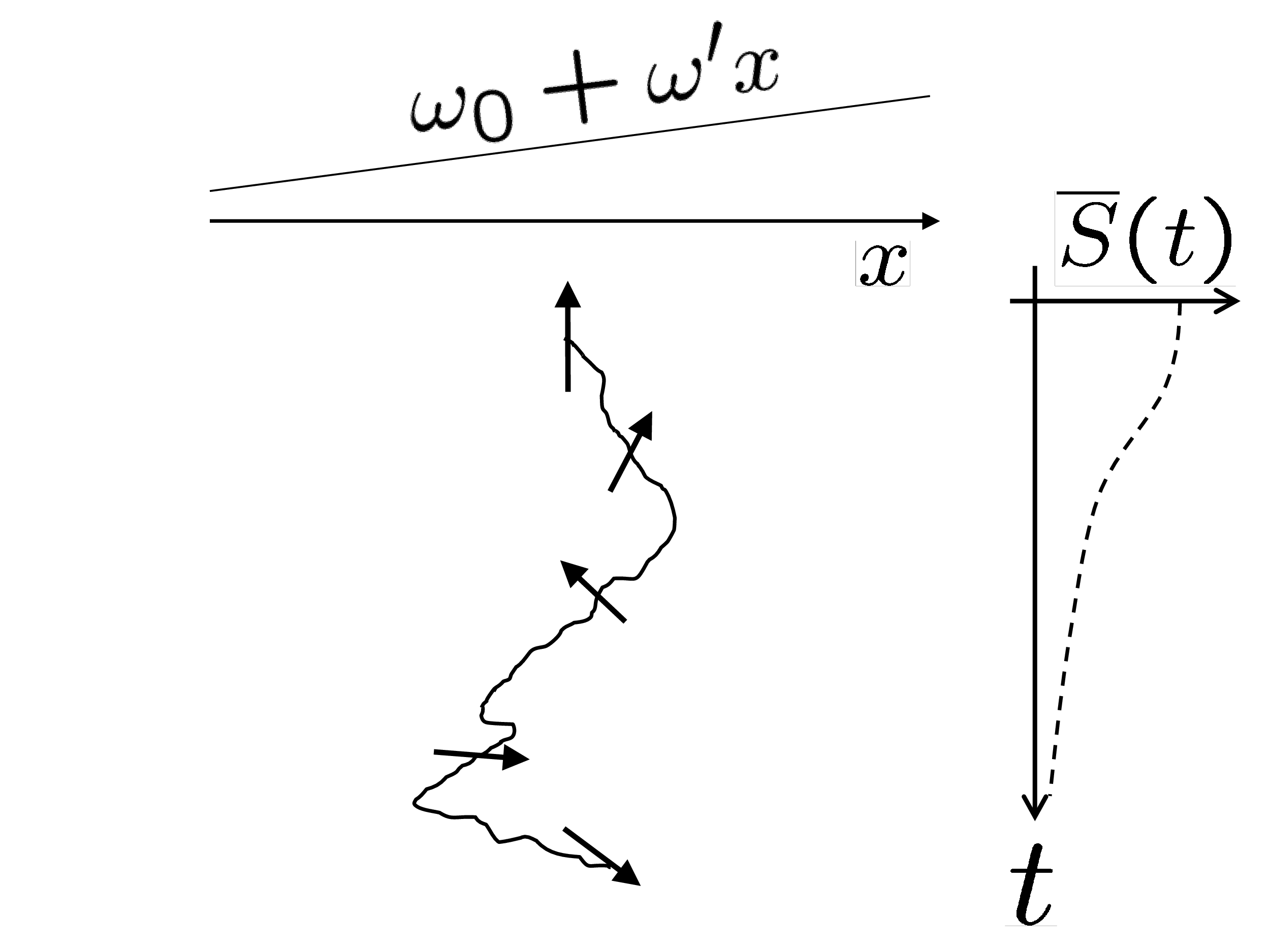,width=0.4\linewidth}}
\caption{The electron performs a random walk. Its spin precesses by
the inhomogeneous magnetic field along the $x$ axis. The average spin
dephases to zero with time.}
\label{fig:inhomogeneous_field}
\end{figure}

\subsection{Quantum mechanical description}

The Born-Markov approximation to the dynamics of a system coupled to an external
environment can be cast in the quantum mechanical language. We refer the reader
to Ref. \cite{Fabian2007:APS} for the derivation. The physics of this derivation
is the same as what we did in Sec. \ref{sec:toy}. Only the formalism is different.

We consider the system described by the Hamiltonian
\begin{equation} \label{eq:toyH}
H(t) = H_0 + V(t),
\end{equation}
in which $H_0$ is our system \emph{per se} and $V(t)$ is the time-dependent
random fluctuating field, of zero average and correlation time $\tau_c$:
\begin{equation}
\overline{V(t)} = \overline{V_(t)} = 0, \quad \overline{V(t) V(t')} \sim e^{-|t-t'|/\tau_c}.
\end{equation}
The system is fully described by the density matrix $\rho$.

The transformation to the rotating frame is equivalent to going to the interaction
picture in quantum mechanics:
\begin{eqnarray}
\rho_I(t) & = & e^{iH_0 t/\hbar} \rho \,  e^{-iH_0 t/\hbar}, \\
V_I(t) & = & e^{iH_0 t/\hbar} V(t) e^{-iH_0 t/\hbar}.
\end{eqnarray}
Performing the operations as outlined in Sec. \ref{sec:toy} for the classical model,
we arrive at the effective time evolution for the density of state operator,
\begin{equation} \label{eq:toy4}
\frac{d \overline{\rho_I(t)}}{d t } = \left (\frac{1}{i\hbar}\right
)^2 \int_0^{t \gg \tau_c} dt' {\left [\overline{V_I(t), \ [V_I(t')},
\overline{\rho_I(t)} ] \right ]}.
\end{equation}
The above equation is called the \emph{Master equation} and is the starting
equation in many important problems in which a quantum system is in contact
with a reservoir.

\subsection{Spin relaxation of conduction electrons}

We will consider nonmagnetic conductors in zero or weak magnetic fields so that the
spin dephasing and spin relaxation times are equal, $T_2 = T_1 = \tau_s$.
The formula for the spin relaxation time,
\begin{equation} \label{eq:motional_narrowing31}
1/\tau_s = \omega^2 \tau_c,
\end{equation}
that was derived above for the spin in a fluctuating magnetic field, applies in
a semiquantitative sense (that is, it gives an order of magnitude estimates and
useful trends) to the conduction electron spins as well. We analyze below the
D'yakonov-Perel'\cite{Dyakonov1972:SPSS} and the Elliott-Yafet\cite{Elliott1954:PR, Yafet:1963}
mechanisms.

\begin{figure}
\centerline{\psfig{file=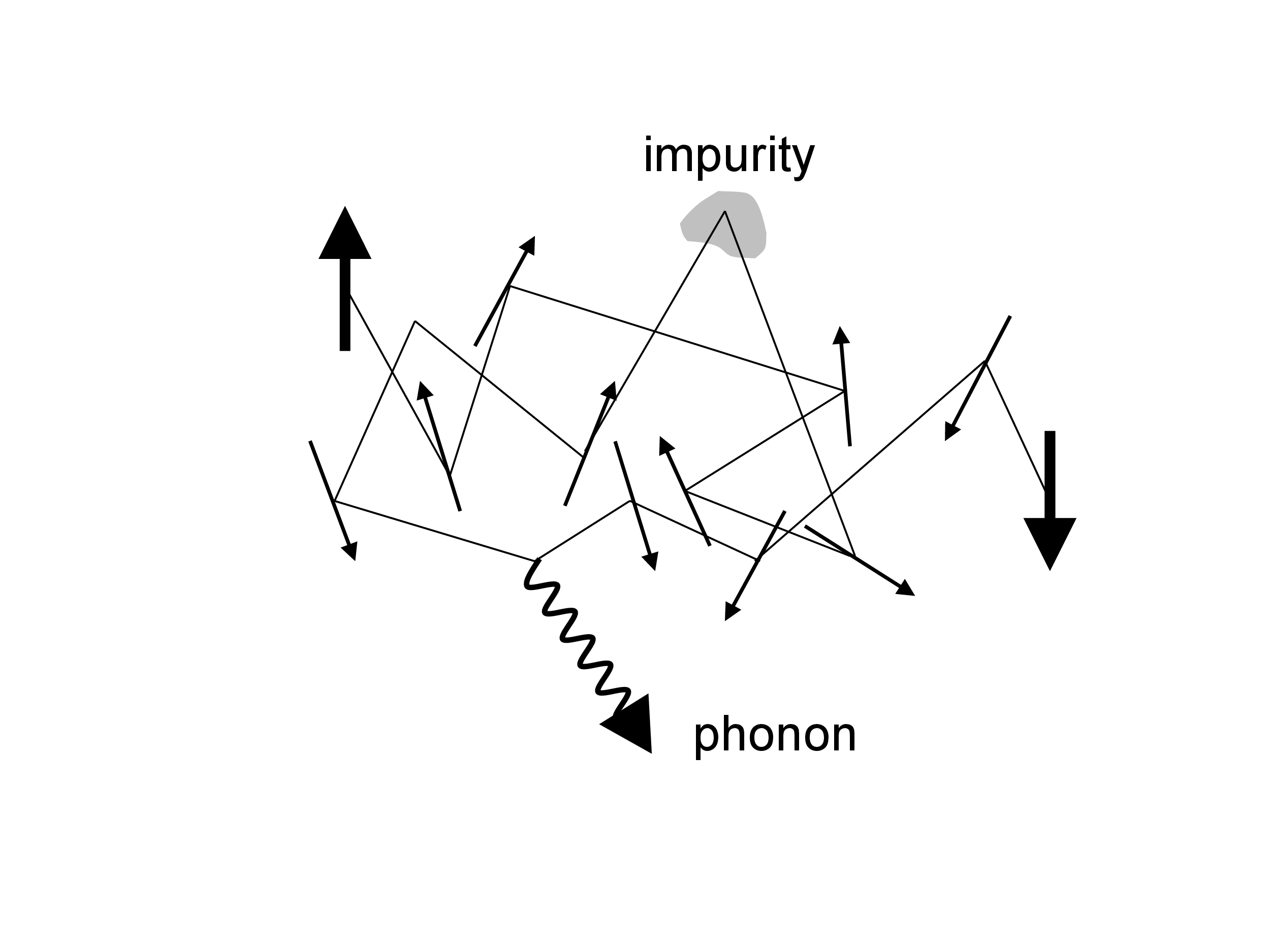,width=0.5\linewidth}}
\caption{{\it D'yakonov-Perel' mechanism.} The electron starts with the spin up. As it moves, its spin precesses
about the axis corresponding to the electron velocity. Phonons and impurities
change the velocity, making the spin to precess about a different axis (and with
different speed). During the scattering event the spin is preserved.}
\label{fig:Dyakonov-Perel}
\end{figure}

The D'yakonov-Perel' mechanism is at play in solids lacking a center of spatial inversion
symmetry. The most prominent example is the semiconductor GaAs. In such solids the
spin-orbit coupling is manifested as some effective magnetic
field---\emph{the spin-orbit field}---that depends on the
electron momentum. Electrons in different momentum states feel different spin-orbit fields, so
that the spin precesses with a given Larmor frequency, until the electron is scattered
into another momentum state. See Fig. \ref{fig:Dyakonov-Perel}.
As the electron momentum changes on the time scale $\tau$ of the momentum relaxation time,
the net effect of the momentum scattering on the spin is to produce random fluctuations
of the Larmor frequencies. We have motional narrowing.
Since these frequencies are correlated by $\tau_c = \tau$, we arrive at
\begin{equation}
\frac{1}{\tau_s} = \omega_{\rm so}^2 \tau,
\end{equation}
for the spin relaxation time. The magnitude of $\omega_{\rm so}$ is the measure of the strength
of the spin-orbit coupling. The spin relaxation rate is directly proportional to the
momentum relaxation time: the more the electron scatters, the less its spin dephases.

\begin{figure}
\centerline{\psfig{file=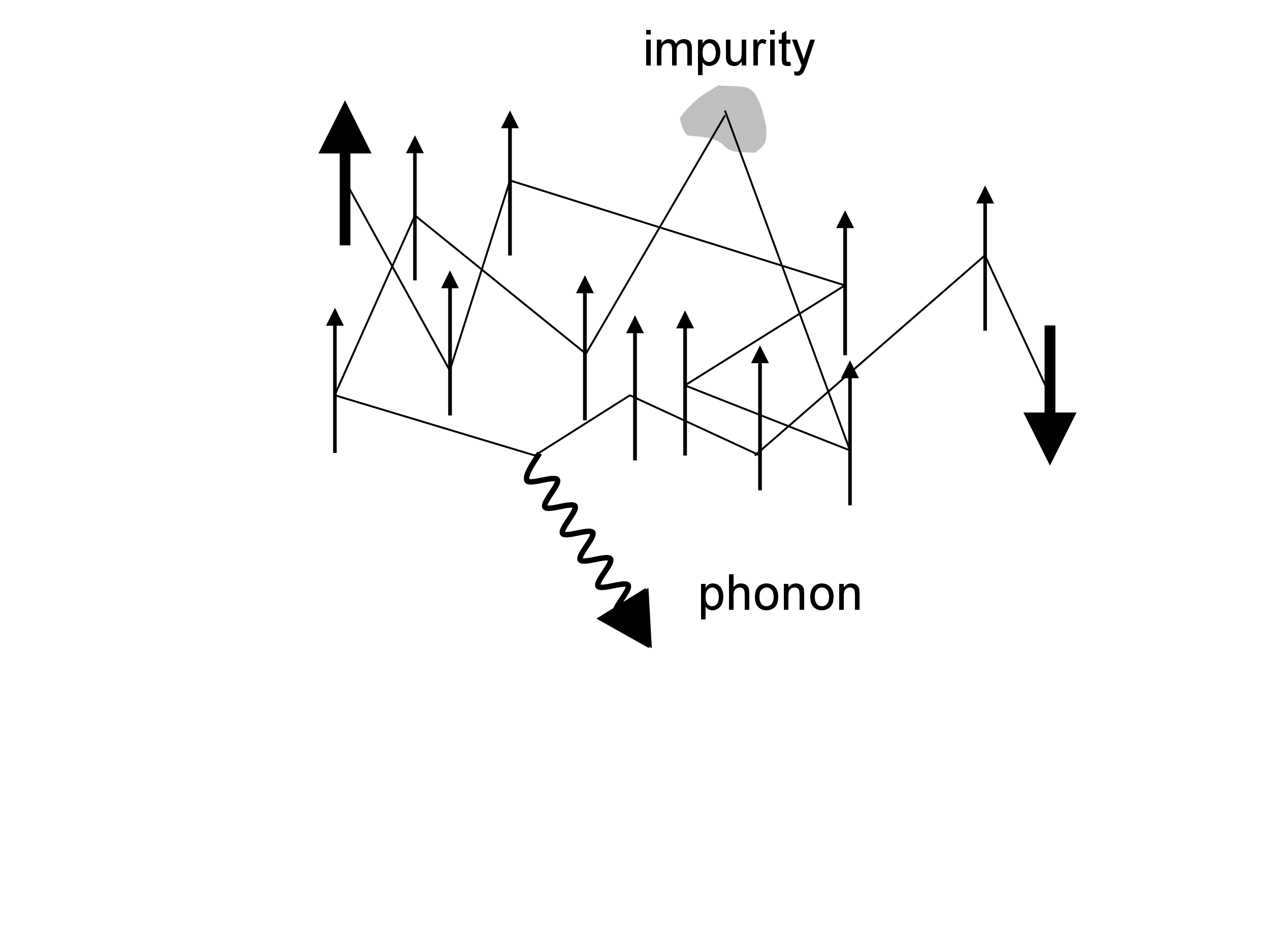,width=0.5\linewidth}}
\caption{{\it Elliott-Yafet mechanism.} The electron starts with the spin up. As it scatters off of
impurities and phonons, the spin can also flip due to spin-orbit coupling.
Between the scattering the spin is preserved.
After, say, a million scattering events, the spin will be down.}
\label{fig:Elliott-Yafet}
\end{figure}

The Elliott-Yafet mechanism works in systems with and without a center of inversion. It relies
on spin-flip momentum scattering. The spin-flip amplitudes are due to spin-orbit coupling, while
the momentum scattering is due to the presence of impurities (that also contribute to the spin-orbit
coupling), phonons, rough boundaries, or whatever is capable of randomizing the electron momentum.
During the scattering events, the spin is preserved. See Fig. \ref{fig:Elliott-Yafet}.
 How do we account for such a scenario with
our toy model? Consider an electron scattering off of an impurity with a spin flip. This spin flip
can be viewed as a precession that occurs during the time of the interaction of the electron
with the impurity. Let us say that the scattering takes the time of $\lambda_F/v_F$, where $v_F$
is the electron velocity and $\lambda_F$ is the electron wave-length at the Fermi level (considering that
it is greater or at most equal to the the size $a$ of the impurity---otherwise we could equally
put $a/v_F$). Then
the precession angle $\varphi = \omega_{\rm so} (\lambda_F/v_F)$, with
$\omega_{so}$ denoting the spin-orbit coupling induced precession frequency. Let us compare
this with the angle of precession, $\omega \tau$, in the motional narrowing model,
Eq. \ref{eq:motional_narrowing31}. We see that
the spin-flip can be described by the effective precession frequency $\omega = \omega_{\rm so} (\lambda_F/v_F\tau)$.
We then obtain for the spin relaxation rate
\begin{equation}
\frac{1}{\tau_s} = \omega_{\rm so}^2 \frac{\lambda_F^2}{v_F^2 \tau}
\approx \left (\frac{\varepsilon_{\rm so}}{\varepsilon_F} \right )^2\frac{1}{\tau}.
\end{equation}
Here $\varepsilon_{\rm so} = \hbar \omega_{\rm so}$ and $\varepsilon= (\hbar k_F) v_F/2$ is the Fermi energy.
For the the Elliott-Yafet mechanism holds that the more the electron scatters, the more the
spin dephases.

Below we discuss the two mechanisms in more detail, providing the formalisms for their
investigation.

\section{The D'yakonov-Perel' mechanism}

D'yakonov and Perel' \cite{Dyakonov1972:SPSS} considered solids without a center of
inversion symmetry, such as GaAs or InAs. The mechanism also works for electrons at
surfaces and interfaces. In such solids the presence of spin-orbit coupling induces
the  spin-orbit fields which give rise to the spin precession. Momentum relaxation
then causes the random walk of the spin phases.

\subsubsection{Spin-orbit field}

In solids without a center of inversion symmetry, spin-orbit coupling splits the electron energies:
\begin{equation}
\varepsilon_{\bf k, \uparrow} \ne \varepsilon_{\bf k, \downarrow}.
\end{equation}
Only the Kramers degeneracy is left, due to time reversal symmetry:
\begin{equation}
\varepsilon_{\bf k, \uparrow} = \varepsilon_{-\bf k, \downarrow}.
\end{equation}
This energy splitting at a given momentum $\bf k$ is conveniently described by the
spin-orbit field $\bgreek{\Omega}$, giving a Zeeman-like (but momentum dependent)
energy contribution to the electronic states, described by the additional Hamiltonian (to the
usual band structure):
\begin{equation}
H_1 = \frac{\hbar}{2} \bgreek{\Omega}_{\bf k}\cdot \bgreek{\sigma}.
\end{equation}
The time reversal symmetry requires that the spin-orbit field is an odd function
of the momentum:
\begin{equation}
\bgreek{\Omega}_{\bf k} = - \bgreek{\Omega}_{-\bf k}.
\end{equation}
The representative example of a spin-orbit field is the Bychkov-Rashba \cite{Bychkov1984:JETPL}
 ($\alpha_{BR}$) and
Dresselhaus \cite{Dresselhaus1955:PR, Dyakonov1986:SPS} ($\gamma_D$) couplings in 2d electron gases formed at the zinc-blende heterostructures
grown along [001] \cite{Fabian2007:APS}:
\begin{equation}
H_1 = (\alpha_{BR} + \gamma_D) \sigma_x k_y - (\alpha_{BR} - \gamma_D) \sigma_y k_x.
\end{equation}
The axes are ${\bf x} = [110]$ and $y = [1\overline{1}0]$. The spin-orbit field
\begin{equation} \label{eq:SOF41}
\hbar{\bgreek{\Omega}}_{\bf k} = 2 \left [(\alpha_{BR} + \gamma_D)k_y , -(\alpha_{BR} - \gamma_D) k_x \right ].
\end{equation}
This field has the $C_{2v}$ symmetry, reflecting the structural symmetry of the zinc-blend
interfaces (such as GaAs/GaAlAs) with the principal axes along [110] and $[1\overline{1}0]$.

\subsubsection{Kinetic equation for the spin}

Let ${\bf s}_{\bf k}$ be the electron spin in the momentum state $\bf k$.
The time evolution of the spin is then described by
\begin{equation} \label{eq:dyakonov-perel41}
\frac{\partial {\bf s}_{\bf k}}{\partial t} - \bgreek{\Omega}_{\bf k} \times {\bf s}_{\bf k}
= - \sum_{{\bf k}'} W_{{\bf k} {\bf k}'} \left ({\bf s}_{\bf k} - {\bf s}_{\bf k}'  \right).
\end{equation}
The left hand side gives the full time derivative $d{\bf s}_{\bf k}/dt$, which is due
to the explicit change of the spin, its direction, and the change of the spin due to the
change of the momentum $\bf k$. The right-hand side is the change of the spin at $\bf k$ due
to the spin-preserving scattering from and to that state. The scattering rate between
the two momentum states $\bf k$ and ${\bf k}'$ is $W_{{\bf k} {\bf k}'}$.

There are two time scales in the problem. First, momentum scattering, which occurs
on the time scale of the momentum relaxation time $\tau$, makes the spins in different
momentum states equal. Second, this uniform spin decays on the time scale of the spin
relaxation time $\tau_s$ which we need to find. We assume that $\tau \ll \tau_s$; this
assumption is well satisfied in real systems. In principle we could go directly to our
model of the electron spin in a fluctuating magnetic field, with the role of the random
Larmor precession playing by $\bgreek{\Omega}$, identifying $\tau_c = \tau$. However,
it is instructive to see how this two time-scale problem is solved directly.

We separate the fast and slow components of the spins,
\begin{equation} \label{eq:fast-slow21}
{\bf s}_{\bf k} = {\bf s} + \bgreek{\xi}_{\bf k}, \quad \langle \bgreek{\xi}_{\bf k} \rangle = 0.
\end{equation}
The symbol $\langle ... \rangle$ denotes averaging over different momenta .
Our goal is to find the effective equation for the time evolution of $\bf s$, which is
the actual spin averaged over $\bf k$. The fast component, $\bgreek{\xi}_{\bf k}$, decays
on  the time scale of $\tau$ to the value given by the instantaneous value of $\bf s$.
Our goal is to find the effective equation for
the time evolution of $\bf s$, which is
the actual spin averaged over $\bf k$. The fast component, $\bgreek{\xi}_{\bf k}$, decays
on  the time scale of $\tau$ to the quasistatic value given by the instantaneous value of $\bf s$.

Upon substituting Eq. \ref{eq:fast-slow21} into the kinetic equation Eq. \ref{eq:dyakonov-perel41}
and averaging over $\bf k$, we obtain the equation of motion for the averaged spin
\begin{equation} \label{eq:dyakonov-perel12}
\dot{\bf s} = \langle \bgreek{\Omega}_{\bf k} \times \bgreek{\xi}_{\bf k},  \rangle
\end{equation}
using that $\langle \Omega_{\bf k} \rangle =0$.
We need to find $\bgreek{\xi}_{\bf k}$. Since $\bf s$ is hardly changing on the
time scales relevant to $\bgreek{\xi}_{\bf k}$, we write,
\begin{equation} \label{eq:dyakonov-perel13}
\dot{\bgreek{\xi}}_{\bf k} - \bgreek{\Omega}_{\bf k} \times {\bf s}  -
\bgreek{\Omega}_{\bf k} \times \bgreek{\xi}_{\bf k} = - \sum_{{\bf k}'}
W_{{\bf k} {\bf k}'} \left (\bgreek{\xi}_{\bf k} - \bgreek{\xi}_{{\bf k}'}   \right ).
\end{equation}
We make the following assumption:
\begin{equation}
\Omega \tau \ll 1.
\end{equation}
That is, we assume that the precession is slow on the time scale of
the momentum relaxation time (see our note on the Born approximation in the toy model
section Sec. \ref{sec:toy}. For simplicity, we make the relaxation time approximation
to model the time evolution of the fast component $\bgreek{\xi}_{\bf k}$:
\begin{equation}
\dot{\bgreek{\xi}}_{\bf k} -  \bgreek{\Omega}_{\bf k} \times {\bf s} -
\bgreek{\Omega}_{\bf k} \times \bgreek{\xi}_{\bf k}
= - \frac{\bgreek{\xi}_{\bf k}}{\tau}.
\end{equation}
From the condition of the quasistatic behavior, $\partial \bgreek{\xi}_{\bf k}/\partial t =0$, \
we get up to the first order in $\Omega \tau$ the following solution for the quasistatic $\bgreek{\xi}_{\bf k}$:
\begin{equation} \label{eq:dyakonov-perel14}
\bgreek{\xi}_{\bf k} = \tau \left (\bgreek{\Omega}_{\bf k} \times {\bf s} \right).
\end{equation}
The above is a realization of coarse graining, in which we effectively average
the spin evolution over the time scales of $\tau$.

Substituting the quasistatic value of $\bgreek{\xi}_{\bf k}$ from Eq. \ref{eq:dyakonov-perel14}
to the time evolution equation for $\bf s$,  Eq. \ref{eq:dyakonov-perel12},  we find
\begin{equation}
\dot{\bf s} = \tau \langle \bgreek{\Omega}_{\bf k} \times \left (\bgreek{\Omega}_{\bf k} \times {\bf s} \right )\rangle.
\end{equation}
Using the vector product identities we finally obtain for the individual spin components $\alpha$,
\begin{equation}
\dot{s}_\alpha = \langle  \Omega_{{\bf k}\alpha} \Omega_{{\bf k} \beta} \rangle \tau s_{\beta}
- \langle \bgreek{\Omega}_{\bf k}^2 \tau \rangle s_\alpha.
\end{equation}
These equations describe the effective time evolution of the electron spin in the presence of
the momentum dependent Larmor precession $\bgreek{\Omega}_{\bf k}$.

For the specific model of the zinc-blend heterostructure with the $C_{2v}$ spin-orbitc
field, Eq. \ref{eq:SOF41}, we obtain the spin dephasing dynamics:
\begin{equation}
\dot{s}_x = - s_x /\tau_s, \quad \dot{s}_y = - s_y /\tau_s, \quad \dot{s}_z = - s_y /\tau_z,
\end{equation}
where
\begin{equation} \label{eq:spin-relaxation61}
\frac{1}{\tau_{x,y}}  = \frac{\left (\alpha_{BR} \mp \gamma_D   \right )^2}{\alpha_{BR}^2 + \gamma_D^2}\frac{1}{\tau_s}, \quad
\frac{1}{\tau_z} = \frac{2}{\tau_s},
\end{equation}
and
\begin{equation}
\frac{1}{\tau_s} = \frac{4 m}{\hbar^4} \varepsilon_k \left ( \alpha_{BR}^2 + \gamma_D^2  \right ).
\end{equation}
The up (down) sign is for the $s_x$ ($s_y$).
The spin relaxation is anisotropic. The  maximum anisotropy is for the case of equal
magnitudes of the Bychkov-Rashba and Dresselhaus interactions, $\alpha_{BR} = \pm \gamma_D$. In this
case one of the spin components does not decay. \footnote{The decay of that component would be
due to higher-order (such as cubic) terms in the spin-orbit fields.} The $s_z$ component of the
spin relaxes roughly twice faster than the in-plane components.

\subsection{The persistent spin helix}

Let us consider the case of $\alpha_{BR} = \gamma_D = \lambda/2$. According to Eq. \ref{eq:spin-relaxation61}
the spin component $s_x$ does not decay, while the decay rates of $s_y$ and $s_z$ are the same:
\begin{equation}
\dot{s}_y = -2 s_y/\tau_s ,\quad \dot{s}_z = - 2 s_z/\tau_s.
\end{equation}
It turns out that a particular nonuniform superposition of $s_y$ and $s_z$ can exhibit no decay as well.
This superposition has been termed \emph{persistent spin helix} \cite{Bernevig2006:PRL}.

Let us assume that the spin is no longer uniform, so that the kinetic equation contains
the spin gradient as well, due to the quasiclassical change of the electronic positions:
\begin{equation}
\frac{\partial {\bf s}_{\bf k}}{\partial t} - \bgreek{\Omega}_{\bf k} \times {\bf s}_{\bf k}
+ \frac{\partial {\bf s}_{\bf k}}{\partial \bf r}\cdot {\bf v}_{\bf k} =
- \sum_{{\bf k}'} W_{{\bf k} {\bf k}'} \left ({\bf s}_{\bf k} - {\bf s}_{\bf k}'  \right).
\end{equation}
A running spin wave,
\begin{equation}
{\bf s}_{\bf k}({\bf r}) = {\bf s}_{\bf k} e^{i {\bf q} \cdot {\bf r}};
\quad {\bf s}_{\bf k} \equiv {\bf s}_{\bf k}({\bf q}),
\end{equation}
then evolves according to,
\begin{equation}
\frac{\partial {\bf s}_{\bf k}}{\partial t} - \bgreek{\Omega}_{\bf k} \times {\bf s}_{\bf k}
+ i \left ( {\bf q} \cdot {\bf v}_{\bf k} \right ) {\bf s}_{\bf k} =
- \sum_{{\bf k}'} W_{{\bf k} {\bf k}'} \left ({\bf s}_{\bf k} - {\bf s}_{\bf k}'  \right).
\end{equation}
We again separate the fast and slow spins,
\begin{equation}
{\bf s}_{\bf k} = {\bf s} + \bgreek{\xi}_{\bf k}, \quad \langle \bgreek{\xi}_{\bf k} \rangle = 0.
\end{equation}
For the dynamics of the slow part we get,
\begin{equation} \label{eq:persistent-spin-helix41}
\dot{\bf s}_{\bf k} = \langle \bgreek{\Omega}_{\bf k} \times \bgreek{\xi}_{\bf k}  \rangle
- i \langle \left ( {\bf q} \cdot {\bf v}_{\bf k} \right )  \bgreek{\xi}_{\bf k}  \rangle.
\end{equation}
Proceeding as in the previous section, assuming that $\bf s$ is stationary on the time
scales relevant to $\bgreek{\xi}$, we can write,
\begin{equation}
\dot{\bgreek{\xi}}_{\bf k} - \bgreek{\Omega}_{\bf k} \times {\bf s}  -
\bgreek{\Omega}_{\bf k} \times \bgreek{\xi}_{\bf k} + i \left ( {\bf q} \cdot {\bf v}_{\bf k}  \right ) {\bf s}
+ i \left ( {\bf q} \cdot {\bf v}_{\bf k}  \right ) {\bgreek{\xi}}_{\bf k} = - \sum_{{\bf k}'}
W_{{\bf k} {\bf k}'} \left (\bgreek{\xi}_{\bf k} - \bgreek{\xi}_{{\bf k}'}   \right ).
\end{equation}
We now work with the following assumptions:
\begin{equation} \label{eq:persistent-spin-helix51}
\Omega \tau \ll 1, \quad q \ell \ll 1,
\end{equation}
where $\ell = g \tau$ is the mean free path. We thus assume that the precession is
slow on the time scale of the momentum relaxation time, as well as (this is new here) the
electronic motion is diffusive on the scale of the wavelength of the spin wave.
In the momentum relaxation approximation, also considering the leading terms according
to the conditions Eq. \ref{eq:persistent-spin-helix51}, we get
\begin{equation}
\dot{\bgreek{\xi}}_{\bf k} -  \bgreek{\Omega}_{\bf k} \times {\bf s} +
i \left ( {\bf q} \cdot {\bf v}_{\bf k}\right ){\bf s} = - \frac{\bgreek{\xi}}{\tau}.
\end{equation}
In the steady-state, corresponding to a given ${\bf s}(t)$, the solution is
\begin{equation}
\bgreek{\xi}_{\bf k} = \tau \left ( \bgreek{\Omega}_{\bf k} \times {\bf s} \right )
-i \tau \left ( {\bf q} \cdot {\bf v}_{\bf k}\right ){\bf s}.
\end{equation}
Substituting to the time evolution equation for $\bf s$, Eq. \ref{eq:persistent-spin-helix51}, we find
\begin{equation}
\dot{\bf s} = - 2 i \tau \langle \left ( {\bf q}\cdot {\bf v}_{\bf k} \right )
\left ( \bgreek{\Omega}_{\bf k} \times {\bf s} \right ) \rangle
+ \tau \langle \bgreek{\Omega}_{\bf k} \times \left (\bgreek{\Omega}_{\bf k} \times {\bf s} \right )\rangle
- \tau \langle \left ( {\bf q} \cdot {\bf v}_{\bf k}  \right )^2  \rangle {\bf s}.
\end{equation}
Using the vector product identities and introducing the diffusivity
\begin{equation}
D = \langle {\bf v}_{{\bf k}\alpha}^2  \rangle \tau,
\end{equation}
where $\alpha$ denote the cartesian coordinates (we assume an isotropic system),
we finally obtain
\begin{equation}
\dot{s}_\alpha = -2 i \tau \varepsilon_{\alpha \beta \gamma} q_{\delta} \langle  v_{{\bf k} \delta} \Omega_{{\bf k} \beta} \rangle s_{\gamma} + \langle  \Omega_{{\bf k}\alpha} \Omega_{{\bf k} \beta} \rangle \tau s_{\beta}
- \langle \bgreek{\Omega}_{\bf k}^2 \tau \rangle s_\alpha - D q^2 s_\alpha.
\end{equation}

For our specific case of
\begin{equation}
\Omega_{\bf k} = (\lambda k_y, 0, 0), \quad \lambda = \alpha_{BR} + \beta_D,
\end{equation}
we find that $s_x$ decays only via diffusion:
\begin{equation}
\dot{s}_x = -D q^2 s_x.
\end{equation}
The spin dephasing is ineffective. This is an expected result.

\begin{figure}
\centerline{\psfig{file=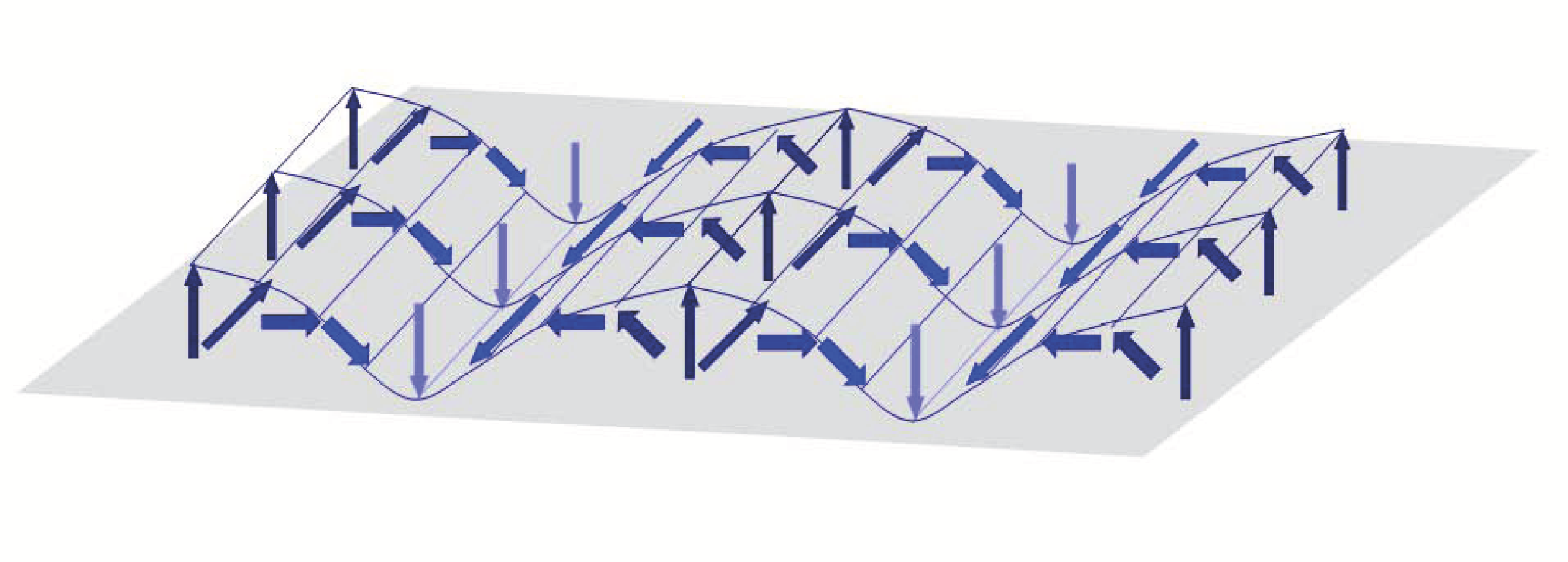,width=0.7\linewidth}}
\caption{The persistent spin helix is a wave of circularly polarized spin. The sense
of polarization, clock or counterclockwise, depends on the relative sign of the Dresselhaus
and the Bychkov-Rashba spin-orbit coupling.}
\label{fig:persistent_spin_helix}
\end{figure}

More interesting behavior is found for the two remaining spin components, $s_y$ and $s_z$.
These two spin components, transverse to the spin-orbit field, are coupled:
\begin{eqnarray}
\dot{s}_y & = & + 2 i \frac{m\lambda}{\hbar} q_y D s_z - \left (\tau \Omega^2 + D q^2   \right ) s_y , \\
\dot{s}_z & = & - 2 i \frac{m\lambda}{\hbar} q_y D s_y - \left (\tau \Omega^2 + D q^2   \right ) s_z ,
\end{eqnarray}
and we denoted $\Omega^2 \equiv \langle \bgreek{\Omega}_{\bf k}^2 \rangle$. Interestingly, this set
of coupled differential equations has only decaying solutions. This is best seen by looking at the
rotating spins:
\begin{equation}
s_+ = s_y + i s_z, \quad s_- = s_y - i s_z,
\end{equation}
whose time evolutions are uncoupled,
\begin{eqnarray}
\dot{s}_+ & = & - \left ( \Omega^2 \tau + Dq^2 - \frac{2 m \lambda}{\hbar} q_y D  \right ) s_+, \\
\dot{s}_- & = & - \left ( \Omega^2 \tau + Dq^2 +  \frac{2 m \lambda}{\hbar} q_y D  \right ) s_-
\end{eqnarray}
Considering that
\begin{equation}
\Omega^2 \tau = \lambda^2 \langle k_y^2 \rangle \tau =
\lambda^2 \left (\frac{m}{\hbar}  \right )^2 \langle v_y^2 \rangle \tau
= \lambda^2  \left (\frac{m}{\hbar}  \right )^2 D,
\end{equation}
we find that the decay of $s_+$ vanishes for the wave vector
\begin{equation}
q_y^{\rm PSH} = \frac{m\lambda}{\hbar} = \frac{m}{\hbar} \left (\alpha_{BR} + \beta_{D} \right ).
\end{equation}
The abbreviation $\rm PSH$ stands for the \emph{persistent spin helix}, which is the spin
wave described by $s_+$ at this particular wave vector. See Fig. \ref{fig:persistent_spin_helix}.
While individually both $s_y$ and $s_z$
decay at a generic wave vector (and also in the uniform case, $q=0$), the spin helix they
form does not decay in the approximation of the linear spin-orbit field. The spin wave rotating
in the opposite sense, $s_-$, on the other hand, decays. And vice versa for
$q_y^{\rm PSH} = - m\lambda/\hbar $. The persistent spin helix was observed in
the spin grating experiment \cite{Koralek2009:N}.

\section{The Elliott-Yafet mechanism}

Elliott \cite{Elliott1954:PR} was first to recognize the role of the intrinsic spin-orbit
coupling---that coming from the host ions---on spin relaxation. Yafet \cite{Yafet:1963}
significanly extended this theory to properly treat electron-phonon spin-flip scattering.
The Elliott-Yafet mechanism dominates the spin relaxation of conduction electrons
in elemental metals and semiconductors, the systems with space inversion symmetry. In
systems lacking this symmetry, the mechanism appears competes with the D'yakonov-Perel'
one; the dominance of one over the other depends on the material in question and
specific conditions, such as temperature and doping.

Suppose a nonequilibrium spin accumulates in a nonmagnetic degenerate conductor.
The spin accumulation is given as the difference between the chemical potentials for the spin up
and the spin down electrons. Let us denote the corresponding potentials by $\mu_\uparrow$ and
$\mu_{\downarrow}$. The nonequilibrium electron occupation function for the spin $\lambda$ is
\begin{equation} \label{eq:EY11}
f_{\lambda {\bf k}} \approx  \frac{1}{e^{\beta(\varepsilon_{\bf k} - \mu_\lambda )} +1} \approx
f^0_{\bf k} + \left [ - \frac{\partial f^0_{\bf k}}{\partial \varepsilon_{\bf k}}
\right ](\mu_\lambda - \varepsilon_F).
\end{equation}
Here,
\begin{equation}
f^0_{\bf k} = f^0(\varepsilon_{\bf k}) =  \frac{1}{e^{\beta(\varepsilon_{\bf k} - \epsilon_F)}+1},
\end{equation}
describes the equilibrium degenerate electronic system of the Fermi energy $\epsilon_F$.
The electron density for the spin $\lambda$ is
\begin{equation}
n_\lambda = \int d\varepsilon g_s(\varepsilon) f^0_{\lambda}(\varepsilon) \approx \frac{n}{2}
+ \int d\varepsilon g(\varepsilon) \left [- \frac{\partial f^0(\epsilon)}{\partial \varepsilon}
  \right ] (\mu_\lambda - \varepsilon_F).
\end{equation}
Here $g_s(\varepsilon)$ is the electronic density of states, defined per unit volume and per spin,
at the fermi level:
\begin{equation}
g_s = \sum_{{\bf k}} \left [ -\frac{\partial f^0_{\bf k}}{\partial \varepsilon_{{\bf k}}} \right ].
\end{equation}
The total electron density $n$ is
\begin{equation}
n = 2 \int d\varepsilon g(\varepsilon) f^0(\varepsilon).
\end{equation}
We assume that the spin accumulation does not charge the system, that is, the charge neutrality
is preserved $n_\uparrow + n_\downarrow = n$. This condition is well satisfied in metals and
degenerate semiconductors that we consider. We then get
\begin{equation}
\mu_\uparrow + \mu_\downarrow = 2 \varepsilon_F.
\end{equation}
The spin density is,
\begin{equation}
s = n_\uparrow - n_\downarrow = g_s (\mu_\uparrow - \mu_\downarrow) = g_s \mu_s,
\end{equation}
where $\mu_s$ is the spin quasichemical potential, $\mu_s = (\mu_\uparrow - \mu_\downarrow)$.

The spin relaxation time $T_1$ is defined by the decay law,
\begin{equation} \label{eq:EY71}
\frac{d s}{d t} = \frac{dn_\uparrow }{dt} - \frac{d n_\downarrow}{dt} = W_{\uparrow \downarrow}
- W_{\downarrow \uparrow} =  -\frac{s}{T_1} = -g_s \frac{\mu_s}{T_1}.
\end{equation}
Here $W_{\uparrow \downarrow}$ is the net number of transitions per unit time from the spin
$\downarrow$ to $\uparrow$. Similarly, $W_{\downarrow \uparrow}$ expresses the rate of spin
flips from $\uparrow$ to $\downarrow$. In the degenerate conductors the spin decay is directly
proportional to the decay of the spin accumulation $\mu_s$:
\begin{equation}
\frac{d \mu_s}{d t} = -\frac{\mu_s}{T_1}.
\end{equation}

\subsection{The electron-impurity scattering}

We need to distinguish the cases of the impurity or host induced spin-orbit
coupling. Altahugh the formulas for the calculation of $T_1$ look similar in the
two cases, they are nevertheless conceptually different.

\subsubsection{Spin-orbit coupling by the impurity}

If the spin-orbit coupling comes from the impurity potential (in this case the
coupling is often termed {\it extrisnic}) the spin-flip scattering is
due to that potential.

The number of transitions per unit time from the spin up to the spin down states is
\begin{equation}
W_{\uparrow\downarrow} = \sum_{{\bf k}n} \sum_{{\bf k}'n'}
W_{{\bf k}'n'\uparrow,{\bf k}n \downarrow}  - W_{{\bf k}n\downarrow, {\bf k}'n' \uparrow}.
\end{equation}
The rate is given by the spin-flip events from down to up minus the ones from up
to down. We use the Fermi golden rule to write out the individual scattering
rates:
\begin{eqnarray}
W_{{\bf k}'n'\uparrow,{\bf k}n \downarrow} & = & \frac{2\pi}{\hbar} f_{{\bf k}n}
\left ( 1 - f_{{\bf k}'n}'  \right) | U_{{\bf k}n \uparrow,{\bf k}' n' \downarrow}|^2
 \delta \left (\varepsilon_{{\bf k}' n'} - \varepsilon_{{\bf k} n} \right ), \\
W_{{\bf k}n\uparrow,{\bf k}'n' \downarrow} & = & \frac{2\pi}{\hbar} f_{{\bf k}'n'}
\left ( 1 - f_{{\bf k}n}  \right) | U_{{\bf k}' n' \downarrow, {\bf k}n \uparrow}|^2
 \delta \left (\varepsilon_{{\bf k}' n'} - \varepsilon_{{\bf k} n} \right ).
\end{eqnarray}
Substituting for the occupation numbers the linearized expression, Eq. \ref{eq:EY11},
and using the definition of the spin relaxation of Eq. \ref{eq:EY71}, we obtain for
the spin relaxation rate the expression,
\begin{equation} \label{eq:EY54}
\frac{1}{T_1} = \frac{2\pi}{\hbar} \frac{1}{g_s} \sum_{{\bf k}{\bf k}'}
 \left |U_{{\bf k}n \uparrow,{\bf k}' n' \downarrow}  \right |^2
 \left [- \frac{\partial f^0(\varepsilon_{\bf k})}{\partial \varepsilon_{\bf k}} \right ]
 \delta \left (\varepsilon_{{\bf k}' n'} - \varepsilon_{{\bf k} n} \right ).
\end{equation}
If we define the spin relaxation for the individual momentum state $\bf k$ as
\begin{equation}
\frac{1}{T_{1 {\bf k}}} = \frac{2\pi}{\hbar} \sum_{{\bf k}'}
 \left |U_{{\bf k}n \uparrow,{\bf k}' n' \downarrow}  \right |^2
 \left [- \frac{\partial f^0(\varepsilon_{\bf k})}{\partial \varepsilon_{\bf k}} \right ]
 \delta \left (\varepsilon_{{\bf k}' n'} - \varepsilon_{{\bf k} n} \right ),
\end{equation}
which has a straightforward interpretation as the spin-flip rate by the
elastic impurity scattering to all the possible states ${\bf k}'$, we get
\begin{equation}
\frac{1}{T_1} = \left  \langle \frac{1}{T_{1{\bf k}}} \right \rangle_{\varepsilon_{\bf k} = \varepsilon_F},
\end{equation}
as the average of the individual scattering rates over the electronic Fermi surface.

\subsubsection{Spin-orbit coupling by the host lattice}

If the spin-orbit coupling comes from the host lattice (in this case the
coupling is often termed {\it intrinsic}) the spin-flip scattering is
due to the admixture of the Pauli spin up and spin down states in the
Bloch eigenstates.

How do the Bloch states actually look like in the presence of spin-orbit
coupling? Elliott showed that the Bloch states
corresponding to a generic lattice wave vector $\bf k$ and band $n$ can be written as
\cite{Elliott1954:PR},
\begin{eqnarray} \label{eq:EY61}
\Psi_{{\bf k}, n \Uparrow} ({\bf r}) &=& \left [a_{{\bf k}n}({\bf
r}) | \uparrow \rangle
+ b_{{\bf k}n} ({\bf r}) |\downarrow \rangle  \right ] e^{i{\bf k}\cdot {\bf r}},  \\
\label{eq:bloch2} \Psi_{{\bf k}, n\Downarrow} ({\bf r}) &=& \left
[a_{-{\bf k}n}^*({\bf r}) | \downarrow \rangle - b_{-{\bf k}n}^*
({\bf r}) |\downarrow \rangle  \right ] e^{i{\bf k}\cdot {\bf r}}. \label{eq:EY62}
\end{eqnarray}
The states $|\uparrow\rangle$ and $|\downarrow\rangle$ are the usual Pauli
spinors. We can select the two states such that $|a_{{\bf k}n}|
\approx 1$ while $|b_{{\bf k}n}| \ll 1$, due to the weak spin orbit
coupling; this justifies calling the two above states ``spin up'' ($\Uparrow$)
and ``spin down'' ($\Downarrow$). In fact, to ``prepare'' the states for the calculation
of the spin relaxation, they need to satisfy,
\begin{equation} \label{eq:EY64}
\langle {\bf k}, n \lambda | \sigma_z | {\bf k}, n \lambda' \rangle = \lambda \delta_{\lambda\lambda'},
\end{equation}
with $\lambda = \Uparrow, \Downarrow$. That is, the two states should
diagonalize the spin matrix $S_z$ (or whatever spin
direction one is interested in).

The Bloch states of Eqs. \ref{eq:EY61} and \ref{eq:EY62} allow for a spin
flip even if the impurity does not induce a spin-orbit coupling. Indeed, the
matrix element
\begin{equation}\label{eq:EY65}
\left \langle {\bf k}, n \Uparrow | U |  {\bf k}, n \Downarrow \right \rangle \sim a b,
\end{equation}
is in general non zero due to the spin admixture. The spin flip probability is
proportional to $|b|^2$, the spin admixture probability. This quantity is
crucial in estimating the spin relaxation in the Elliott-Yafet mechanism.
The spin relaxation time $T_1$ in this case can be calculated using the formula
Eq. \ref{eq:EY54}, with
\begin{equation}
U_{{\bf k}n \uparrow,{\bf k}' n' \downarrow} = U_{{\bf k}n \Uparrow,{\bf k}' n' \Downarrow},
\end{equation}
given by Eq. \ref{eq:EY65}. A useful \emph{rule of thumb} for estimating the spin relaxation
time in this case is
\begin{equation}
\frac{1}{T_1} \approx \frac{\langle b_{{\bf k}n}^2 \rangle }{\tau_{p}},
\end{equation}
where the averaging of the spin admixture probabilities $b^2$ is performed over
the Fermi surface (or the relevant energy scales of the problem); $\tau_p$ is the
spin-conserving momentum relaxation time. We stress that $b$ is obtained from
the states prepared according  to Eq. \ref{eq:EY64}.

\subsection{The electron-phonon scattering}

The spin-flip due to the scattering of the electrons off of phonons involves the
intrinsic spin-orbit potential. The electron Bloch states are the ones
given by the Eqs. \ref{eq:EY61} and \ref{eq:EY62}, prepared according
to Eq. \ref{eq:EY64}.

The net number of transitions per unit time from the spin up to the spin down states is
\begin{equation}
W_{\uparrow\downarrow} = \sum_{{\bf k}n} \sum_{{\bf k}'n'} \sum_{{\bf q} \nu}
W_{{\bf k}n\uparrow, {\bf q} \nu; {\bf k}'n' \downarrow} +
W_{{\bf k}n\uparrow; {\bf k}'n' \downarrow, {\bf q} \nu} -
W_{{\bf k}'n'\downarrow, {\bf q} \nu; {\bf k}n \uparrow} -
W_{{\bf k}'n'\downarrow; {\bf k}n \uparrow, {\bf q} \nu}.
\end{equation}
We introduced the rates of the spin flip transitions accompanied by the phonon
absorption and emissions as follows. The net transition rate from the
single electron state $|{\bf k}' n' \downarrow \rangle$ to the electron state
$|{\bf k} n \uparrow \rangle$ while the phonon of momentum ${\bf q}$ and
polarization $\nu$ is emitted, is
\begin{equation}
W_{{\bf k}n\uparrow, {\bf q} \nu; {\bf k}'n' \downarrow} =
\frac{2\pi}{\hbar} \left | M_{{\bf k} n \uparrow, {\bf q} \nu; {\bf k}' n' \downarrow } \right |^2
f_{{\bf k}' n' \downarrow} \left (1 - f_{{\bf k} n \uparrow}  \right)
\delta \left (\varepsilon_{{\bf k} n} - \varepsilon_{{\bf k}' n'} + \hbar \omega_{{\bf q} \nu}   \right ).
\end{equation}
Similarly, the net transition rate from the
single electron state $|{\bf k}' n' \downarrow \rangle$ to the electron state
$|{\bf k} n \uparrow \rangle$ while the phonon of momentum ${\bf q}$ and
polarization $\nu$ is absorbed, is
\begin{equation}
W_{{\bf k}n\uparrow; {\bf k}'n' \downarrow, {\bf q} \nu} =
\frac{2\pi}{\hbar} \left | M_{{\bf k} n \uparrow; {\bf k}' n' \downarrow {\bf q} \nu } \right |^2
f_{{\bf k}' n' \downarrow} \left (1 - f_{{\bf k} n \uparrow}  \right)
\delta \left (\varepsilon_{{\bf k} n} - \varepsilon_{{\bf k}' n'} - \hbar \omega_{{\bf q} \nu}   \right ).
\end{equation}
The same way are defined the remaining two rates, $W_{{\bf k}'n'\downarrow, {\bf q} \nu; {\bf k}n \uparrow}$
for the spin flip from ${\bf k} n \uparrow$ to ${\bf k}' n' \downarrow$ with the phonon emission,
and  $W_{{\bf k}'n'\downarrow; {\bf k}n \uparrow {\bf q} \nu}$, for the phonon absorption:
\begin{eqnarray}
W_{{\bf k}'n'\downarrow, {\bf q} \nu; {\bf k}n \uparrow} &= & \frac{2\pi}{\hbar} \left | M_{{\bf k}' n' \downarrow, {\bf q} \nu; {\bf k} n \uparrow } \right |^2
f_{{\bf k} n \uparrow} \left (1 - f_{{\bf k}' n' \downarrow}  \right)
\delta \left (\varepsilon_{{\bf k}' n'} - \varepsilon_{{\bf k} n} + \hbar \omega_{{\bf q} \nu}   \right ), \\
W_{{\bf k}'n'\downarrow; {\bf k}n \uparrow {\bf q} \nu} & = &
\frac{2\pi}{\hbar} \left | M_{{\bf k}' n' \downarrow, {\bf q} \nu; {\bf k} n \uparrow,  {\bf q} \nu } \right |^2
f_{{\bf k} n \uparrow} \left (1 - f_{{\bf k}' n' \downarrow}  \right)
\delta \left (\varepsilon_{{\bf k}' n'} - \varepsilon_{{\bf k} n} - \hbar \omega_{{\bf q} \nu}   \right ).
\end{eqnarray}

The calculation of the spin relaxation due to the electron-phonon scattering
is rather involved and we cite here only the final result for degenerate conductors:
\begin{eqnarray} \label{eq:EY100}
\frac{1}{T_1} = & &\frac{4\pi}{\hbar} \frac{1}{g_s} \sum_{{\bf k}n} \sum_{{\bf k}'n'}
\sum_{\nu} \left ( -\frac{\partial f^0_{{\bf k}n}}{\varepsilon_{{\bf k}n}} \right ) \frac{\hbar}{2NM}
\frac{N^2}{\omega_{{\bf k}'- {\bf k}, \nu}} \left | \bgreek{\epsilon}_{{\bf k} - {\bf k}', \nu}
\cdot \langle {\bf k} n \Uparrow \left | \bgreek{\nabla} V  \right | {\bf k}' n' \Downarrow \rangle   \right |^2 \\
&&\times  \left  \{  \left [n_{{\bf q}\nu} - f^0_{{\bf k}'n'} + 1  \right ]
\delta \left ( \varepsilon_{{\bf k}n} - \varepsilon_{{\bf k}'n'} - \hbar \omega_{{\bf q} \nu}  \right )
+   \left [n_{{\bf q}\nu} + f^0_{{\bf k}'n'} \right ]
\delta \left ( \varepsilon_{{\bf k}n} - \varepsilon_{{\bf k}'n'} + \hbar \omega_{{\bf q} \nu}  \right )
  \right \}.
\end{eqnarray}
The electronic bands are described by the energies $\varepsilon_{{\bf k}n}$ of the state with
momentum $\bf k$ and band index $n$. Phonon frequencies are $\omega_{{\bf q}\nu}$, for the phonon
of momentum $\bf q$ and polarization $\nu$; similarly for the phonon polarization vector
$\epsilon_{{\bf q}\nu}$. We further denoted by $M$ the atomic mass, by $N$ the number of atoms
in the lattice, and by $\bgreek{\nabla} V$ the gradient of the electron-lattice ion potential.
The equilibrium phonon occupation numbers are denoted as $n_{{\bf q} \nu}$, given
as
\begin{equation}
n_{{\bf q}\nu} = n(\omega_{{\bf q} \nu}) = \frac{1}{e^{\beta \hbar \omega_{{\bf q}\nu}} -1}.
\end{equation}
The electronic states $|{\bf k} n \sigma \rangle$ are normalized to the whole space.

Two types of processes contribute to the phonon induced spin flips. First, what we call
the \emph{Elliott processes}, are the Elliott-type of spin flips in which the Bloch states
given by Eq. \ref{eq:EY61} and \ref{eq:EY62} scatter by the scalar part of the gradient of the
electron-ion potential $V$. Second, what we call the \emph{Yafet processes}, are the
spin flips due to the gradient of the spin-orbit part of the electron-ion potential.
These two processes are typically of similar order of magnitude and have to be added
coherently in order to obtain $T_1$.

\begin{figure}
\centerline{\psfig{file=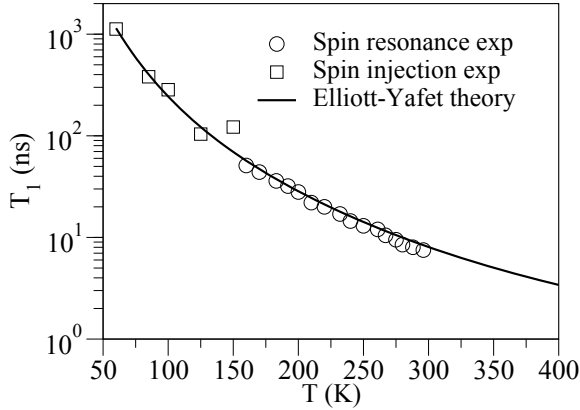,width=0.5\linewidth}}
\caption{\emph{Spin relaxation in silicon.} Phonon-induced spin relaxation in
silicon results in an approximate $T^3$ power law \cite{Cheng2010:PRL}. Shown are the experimental
data from spin resonance \cite{Lepine1970:PRB} and spin injection \cite{Huang2007:PRL} experiments,
and a calculation based on the Elliott-Yafet mechanism \cite{Cheng2010:PRL}.}
\label{fig:spin_relaxation_silicon}
\end{figure}

Figure \ref{fig:spin_relaxation_silicon} shows the experimental data of the spin relaxation in
intrinsic (nondegenerate) silicon, obtained by the spin resonance \cite{Lepine1970:PRB, Fabian2007:APS} and the spin injection
 \cite{Huang2007:PRL} experiments. The calculation based on the Elliott-Yafet mechanism of
 the phonon induced spin flips reproduces the experiments \cite{Cheng2010:PRL}.

\subsubsection{The Yafet relation}

The expected temperature dependence of the phonon-induced spin flips in degenerate conductors, is, according to
the Elliott-Yafet mechanism, $1/T_1 \sim T$ at high temperatures (starting roughly at a fraction of the Debye temperature $T_D$) and
$1/T_1 \sim T^3$ at low temperatures, in analogy with the conventional spin-conserving electron-phonon
scattering. The high-temperature dependence originates from the linear increase of the phonon occupation numbers
$n$ with increasing temperature: $n_{{\bf q}\nu} \sim T$, as $k_B T \gg \omega_{{\bf q} \nu}$. At $T > T_D$  are
all the phonons excited. The low temperature dependence of the spin-conserving scattering follows from setting
the relevant phonon energy scale to $k_B T$. The matrix element
\begin{equation}
\left \langle {\bf k} n \Uparrow \left | \bgreek{\nabla} V  \right | {\bf k}' n' \Uparrow \right \rangle \sim q,
\end{equation}
which gives the $1/T_1 \sim T^3$ dependence. However, Yafet showed \cite{Yafet:1963} that for the
spin-flip matrix element the space inversion symmetry modifies the momentum dependence to
\begin{equation} \label{eq:EY77}
\left \langle {\bf k} n \Uparrow \left | \bgreek{\nabla} V  \right | {\bf k}' n' \Downarrow \right \rangle \sim q^2,
\end{equation}
so that
\begin{equation}
\frac{1}{T_1} \sim T^5,
\end{equation}
instead of the expected $T^5$.
Since the same temperature dependence holds for the phonon-induced electrical resistance $\rho(T)$, we can write
\begin{equation}
\frac{1}{T_1} \sim \rho(T),
\end{equation}
known as the \emph{Yafet relation}.

\begin{figure}
\centerline{\psfig{file=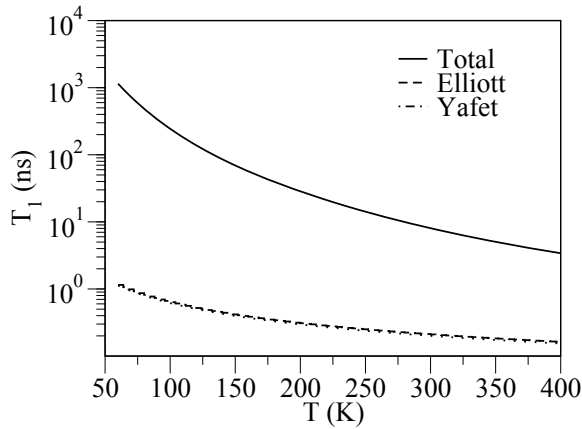,width=0.5\linewidth}}
\caption{\emph{Interference between the Elliott and Yafet processes.} Individually the
Elliott and Yafet phonon-induced spin relaxation processes give spin relaxation
orders of magnitude stronger that the total one. This example is from the calculation
of $T_1$ in silicon \cite{Cheng2010:PRL}. }
\label{fig:elliott-yafet-interference}
\end{figure}

The relation Eq. \ref{eq:EY77} holds if both the Elliott and Yafet processes are taken into account. Individually, they
would lead to a linear dependence on $q$. The quantum  mechanical interference between these two processes
thus significantly reduces the spin-flip probability at low momenta $q$. An example is shown in Fig.
\ref{fig:elliott-yafet-interference}. The Elliott and Yafet processes would individually give much stronger
spin relaxation than is observed. Their destructive interference can modify $T_1$ by orders of magnitude.

\section{Results based on kinetic spin Bloch equation approach}

It was shown by Wu {\em et al}. from a full microscopic
kinetic-spin-Bloch-equation  approach \cite{wu-review}
 that the single-particle
approach is inadequate in accounting for the spin relaxation/dephasing
both in the time \cite{wu, wu2, wu3, weng, weng2}
 and in the space \cite{weng3, weng4, weng5}
domains. The momentum
dependence of the effective magnetic field (the D'yakonov-Perel' term) and the
momentum dependence of the spin diffusion rate along the spacial
gradient \cite{weng3} or even the random spin-orbit coupling \cite{ya} all
serve as inhomogeneous broadenings \cite{wu2, wu3}. It was pointed out that in
the presence of inhomogeneous broadening, any scattering, including
the carrier-carrier Coulomb scattering, can cause an irreversible spin
relaxation/dephasing \cite{wu2}. Moreover, besides the spin relaxation/dephasing channel the scattering
provides, it also gives rise to the counter effect to the
inhomogeneous broadening. The scattering tends to drive carriers to a
more homogeneous state and therefore suppresses the inhomogeneous
broadening. Finally,  this approach is valid in both strong and weak
scattering limits and also  can be used to study systems far away from the
equilibrium, thanks to the inclusion of the Coulomb
scattering.

In the following, we present the main results based on
kinetic-spin-Bloch-equation  approach. We first briefly introduce
the kinetic spin Bloch equations. Then  we review the results of the
spin relaxation/dephasing in the time and space domains respectively.

\subsection{Kinetic spin Bloch equations}

By using the nonequilibrium Green function method with gradient
expression as well as the generalized Kadanoff-Baym Ansatz \cite{haug}, we
construct the kinetic spin Bloch equations as follows:
\begin{eqnarray}
\dot \rho_{\mathbf{k}}(\mathbf{r}, t) &=&
\left.\dot\rho_{\mathbf k}(\mathbf{r}, t)
\right|_{\mathtt{dr}}
+ \left. \dot \rho_{\mathbf k}(\mathbf{r}, t)
\right|_{\mathtt{dif}}\nonumber\\
&&+ \left. \dot \rho_{\mathbf k}(\mathbf{r}, t)
\right|_{\mathtt{coh}}
+ \left. \dot\rho_{\mathbf k}(\mathbf{r},t)\right|_{\mathtt{scat}}\ .
\label{KSBE}
\end{eqnarray}
Here $\rho_{\mathbf k}(\mathbf r, t)$
are the density matrices of electrons
with momentum ${\bf k}$ at position ${\bf r}$ and time $t$.
The off-diagonal elements of $\rho_{\mathbf  k}$
 represent the correlations between the conduction and valence bands,
different subbands (in confined structures) and different spin states.
$\left.\dot\rho_{\mathbf k}(\mathbf{r}, t)\right|_{\mathtt{dr}}$
are the driving terms from the external electric field.
The coherent terms in Eq.\ (\ref{KSBE})
$\dot\rho_{\bf k}|_{\mathtt{coh}}$ are composed of the energy spectrum, magnetic
field and effective magnetic field from the D'yakonov-Perel' term, and the Coulomb Hartree-Fock
terms. The diffusion terms
$\left. \dot \rho_{\mathbf k}(\mathbf{r}, t)
\right|_{\mathtt{dif}}$ come from the spacial gradient.
The scattering terms $ \left. \dot\rho_{\mathbf k}(\mathbf{r},t)\right|_{\mathtt{scat}}$
include the spin-flip and spin conserving electron-electron, electron-phonon and
electron-impurity scatterings. The spin-flip terms correspond to the
Elliot-Yafet and/or Bir-Aronov-Pikus mechanisms. Detailed expressions of these terms in the kinetic spin Bloch equations
depend on the band structures, doping situations and dimensionalities
\cite{wu-review} and can be found in the literature for different cases, such as
intrinsic quantum wells \cite{wu}, $n$-type quantum wells
without \cite{weng,weng6,zhou1,zhou2,jiang1} and
with \cite{weng2,zhang1,jiang2,zhang2} electric field,
$p$-type quantum wells \cite{lu,yzhou,zhang3}, quantum wires \cite{lu1,lu2},
quantum dots \cite{jiang3} and bulk materials \cite{jiang4} in the spacial
uniform case and quantum wells in spacial non-uniform
case \cite{weng3,weng4,cheng,zhang4}.
 By numerically solving the kinetic spin Bloch equations
 with all the scattering explicitly
included, one is able to obtain the time evolution and/or
spacial distribution of the density matrices,
and hence all the measurable quantities, such as mobility,
diffusion constant, optical relaxation/dephasing time, spin relaxation/dephasing time, spin diffusion length,
as well as hot-electron temperature, can be
determined from the theory without any fitting parameters.

\subsection{Spin relaxation/dephasing}

In this subsection we present the main understandings of the
spin relaxation/dephasing  added to the literature
from the kinetic-spin-Bloch-equation approach.  We focus on three related
issues: (i) The importance of
the Coulomb interaction to the spin relaxation/dephasing; (ii) Spin dynamics far away
from the equilibrium; and (iii) Qualitatively
different behaviors from those wildly adopted in the literature.

First we address the effect of the Coulomb interaction.
Based on the single-particle approach, it has been long believed
that the Coulomb scattering is irrelevant to the spin relaxation/dephasing
\cite{flat}.  It was first pointed out by Wu
and Ning \cite{wu2} that in the presence of inhomogeneous broadening
in spin precession, i.e., the spin precession frequencies are ${\bf
  k}$-dependent, any scattering, including the spin-conserving
scattering, can cause irreversible spin dephasing. This inhomogeneous
broadening can come from the energy-dependent
$g$-factor \cite{wu2}, the D'yakonov-Perel' term \cite{wu3}, the
 random spin-orbit coupling \cite{ya}, and even the
momentum dependence of the spin diffusion rate along the spacial
gradient \cite{weng3}.
Wu and Ning first showed that with the energy-dependent $g$-factor as
an inhomogeneous broadening, the Coulomb scattering can lead to
irreversible spin dephasing \cite{wu2}. In [001]-grown $n$-doped
quantum wells, the importance of the Coulomb scattering for spin
relaxation/dephasing was proved by Glazov and Ivchenko \cite{glazov}
by using perturbation theory and by Weng and Wu \cite{weng}
from the kinetic-spin-Bloch-equation  approach. In a
temperature-dependent study of the spin dephasing in [001]-oriented
$n$-doped quantum wells, Leyland et al. experimentally verified
the effects of the electron-electron Coulomb scattering by
closely measuring the momentum scattering rate from the mobility \cite{brand}.
By showing the momentum relaxation rate obtained from the mobility cannot
give the correct spin relaxation rate, they showed the difference comes
from the Coulomb scattering.  Later
Zhou et al. even predicted a peak from the Coulomb scattering in
the temperature dependence of the spin relaxation time in a
high-mobility low-density $n$-doped (001)
quantum well \cite{zhou1}. This was later demonstrated by Ruan et al.
experimentally \cite{ruan}.

\begin{figure}[htb]
 \begin{center}
 \includegraphics[width=6cm]{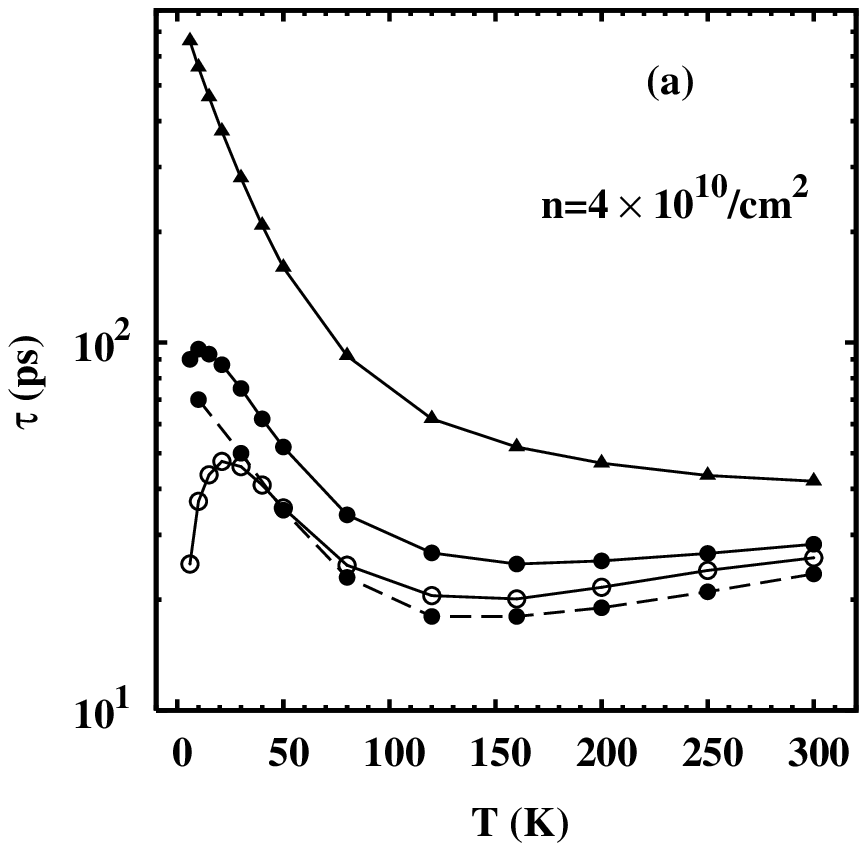}
\includegraphics[width=6cm]{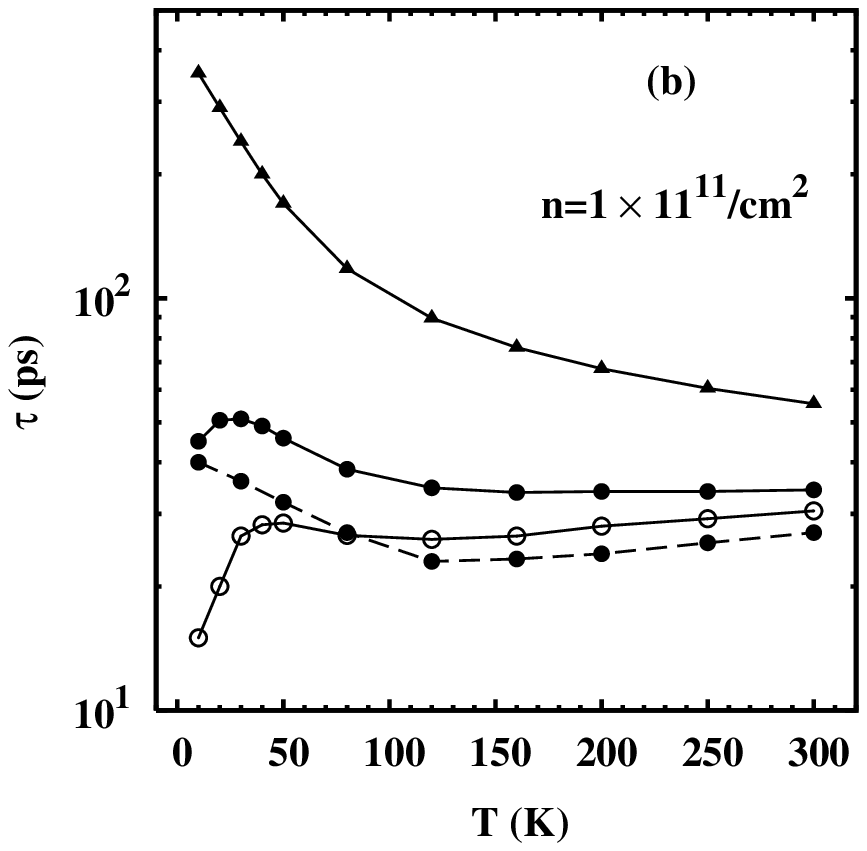}
\includegraphics[width=6cm]{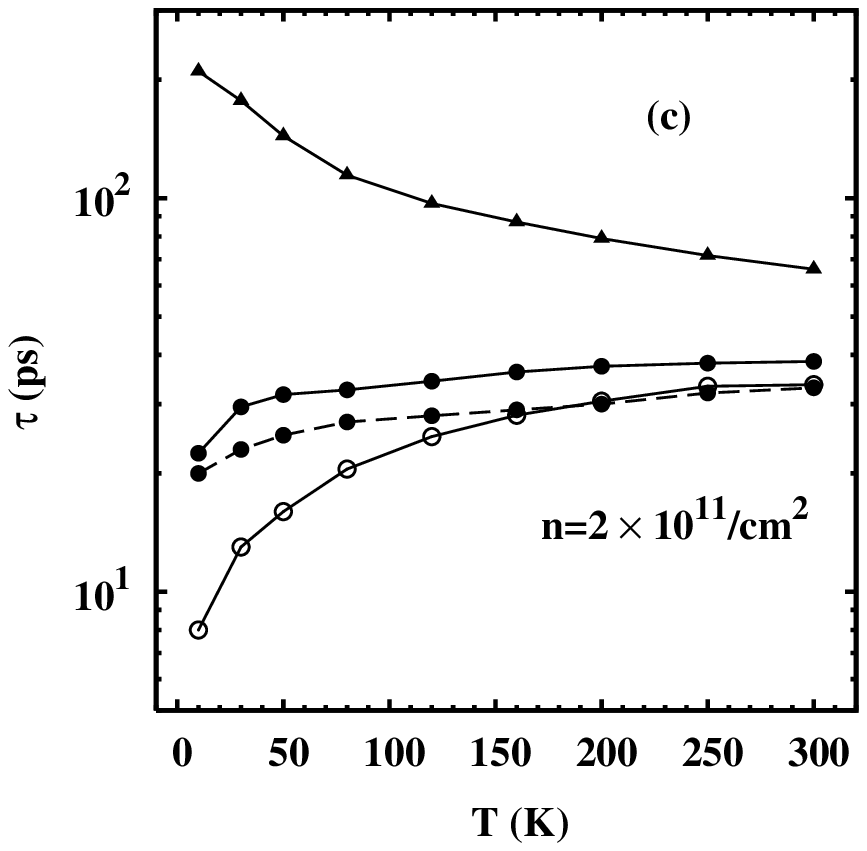}
\end{center}
 \caption{Spin relaxation time $\tau$ vs. the temperature $T$ with well
   width $a=7.5$~nm and electron density $n$ being (a) $4\times
   10^{10}$~cm$^{-2}$, (b) $1\times 10^{11}$~cm$^{-2}$, and (c) $2\times
   10^{11}$~cm$^{-2}$, respectively. Solid curves with triangles:
   impurity density $n_i=n$; solid curves with dots: $n_i=0.1n$; solid
   curves with circles: $n_i=0$; dashed curves with dots: $n_i=0.1n$ and
   no Coulomb scattering. From Zhou et al. \cite{zhou1}.}
 \label{fig1}
 \end{figure}

Figure~\ref{fig1} shows the temperature dependence of the spin
relaxation time of a 7.5~nm GaAs/Al$_{0.4}$Ga$_{0.6}$As
quantum well at different electron and
impurity densities \cite{zhou1}. For this small well width, only
the lowest subband is needed in the calculation. It is shown in the
figure that when the electron-impurity scattering is dominant, the
spin relaxation time decreases with
increasing temperature monotonically. This is
in good agreement with the experimental findings \cite{ohno}
and a nice agreement of the theory and the experimental data from 20 to 300~K
is given in Ref. \cite{zhou1}.
However, it is shown that for
sample with high mobility, i.e., low impurity density, when the
electron density is low enough, there is a peak at low
temperature. This peak, located around the Fermi temperature of
electrons $T_F^e=E_F/k_B$, is identified to be solely due to the Coulomb
scattering \cite{zhou1,bronold}. It disappears when the Coulomb
scattering is switched off, as shown by the dashed curves in the
figure. This peak also disappears at high impurity densities.
It is also noted in
Fig.~\ref{fig1}(c) that for electrons of high density so that
$T_F^e$ is high enough and the contribution from the
electron--longitudinal optical-phonon
scattering becomes marked, the peak disappears even for sample with no
impurity and the spin relaxation time increases with temperature
monotonically. The physics leading to the peak is due to the crossover
of the Coulomb scattering from the degenerate to the non-degenerate
limit. At $T<T_F^e$, electrons are in the degenerate limit  and the
electron-electron scattering rate $1/\tau_{\rm ee}\propto T^{2}$. At
$T>T_F^e$, $1/\tau_{\rm ee}\propto T^{-1}$ \cite{glazov,gio}. Therefore, at
low electron density so that $T_F^e$ is low enough and the electron-acoustic
phonon scattering is very weak comparing with the electron-electron
Coulomb scattering, the Coulomb scattering is dominant
for high mobility sample. Hence the different temperature dependence
of the Coulomb scattering leads to the peak. It is noted that the peak
is just a feature of the crossover from the degenerate to the
non-degenerate limit. The location of the peak also depends on the
strength of the inhomogeneous broadening. When the inhomogeneous
broadening depends on momentum linearly, the peak tends to appear at
the Fermi temperature. A similar peak was predicted in the electron spin
relaxation in $p$-type GaAs quantum well and the hole spin
relaxation in (001) strained asymmetric Si/SiGe quantum well, where
the electron and hole spin relaxation times both show a peak at the
hole Fermi temperature $T_F^h$ \cite{zhang3,yzhou}. When the
inhomogeneous broadening depends on momentum cubically, the peak tends
to shift to a lower temperature. It was predicted that a peak in the
temperature dependence of the electron spin relaxation time appears at
a temperature in the range of ($T_F^e/4$, $T_F^e/2$) in the intrinsic
bulk GaAs \cite{jiang4} and a peak in the temperature dependence
of the hole spin relaxation time at $T_F^h/2$ in $p$-type Ge/SiGe
quantum well \cite{zhang3}.
 Ruan {\em et al}. demonstrated the
peak experimentally in a high-mobility low-density
GaAs/Al$_{0.35}$Ga$_{0.65}$As heterostructure
and showed a peak appears
at $T_F^e/2$ in the spin relaxation time versus temperature
curve \cite{ruan}.

For larger well width, the situation may become different
in the non-degenerate limit. Weng and Wu calculated the spin
relaxation/dephasing for (001) GaAs quantum wells with
larger well width and high mobility, by including the multi-subband
effect \cite{weng6}. It is shown that for small/large well width so that the
linear/cubic Dresselhaus term is dominant, the spin relaxation/dephasing time
increases/decreases with the temperature. This is because with the
increase of temperature, both the inhomogeneous broadening and the
scattering get enhanced. The relative importance of these two
competing effects is different when the linear/cubic term is
dominant \cite{weng6}.  Jiang and
Wu further introduced strain to change the relative importance of the
linear and cubic D'yakonov-Perel' terms and showed the different temperature
dependences of the spin relaxation time \cite{ljiang}. This
prediction has been realized experimentally by Holleitner {\em et al.}
 where they showed that in $n$-type two-dimensional InGaAs
channels, when the linear D'yakonov-Perel' term is suppressed, the spin relaxation
time decreases with temperature monotonically \cite{holl}. Another interesting
prediction related to the multi-subband effect is related
to the effect of the inter-subband Coulomb scattering. From the
calculation Weng and Wu found out
that although the inhomogeneous broadening from the higher subband
of the (001) quantum well is much larger, due to the strong inter-subband
Coulomb scattering, the spin relaxation times of the lowest two subbands are
identical \cite{weng6}. This prediction has later been verified experimentally
by Zhang {\em et al.},  who
studied the spin dynamics in a single-barrier heterostructure by
time-resolved Kerr rotation \cite{zheng2}. By applying a gate voltage, they
effectively manipulated the confinement of the second subband and the
measured spin relaxation times of the first and second subbands are
almost identical at large gate voltage. L\"u {\em et al.}
showed that due to the Coulomb scattering, $T_2=T_2^\ast$ in (001) GaAs
quantum wells for a wide temperature and density regime \cite{lu3}.
It was also pointed out by L\"u {\em et al.} that in the strong (weak)
scattering  limit, introducing the Coulomb scattering will always lead to
a faster (slower) spin relaxation/dephasing \cite{lu}.

\begin{figure}[htb]
\centerline{\psfig{file=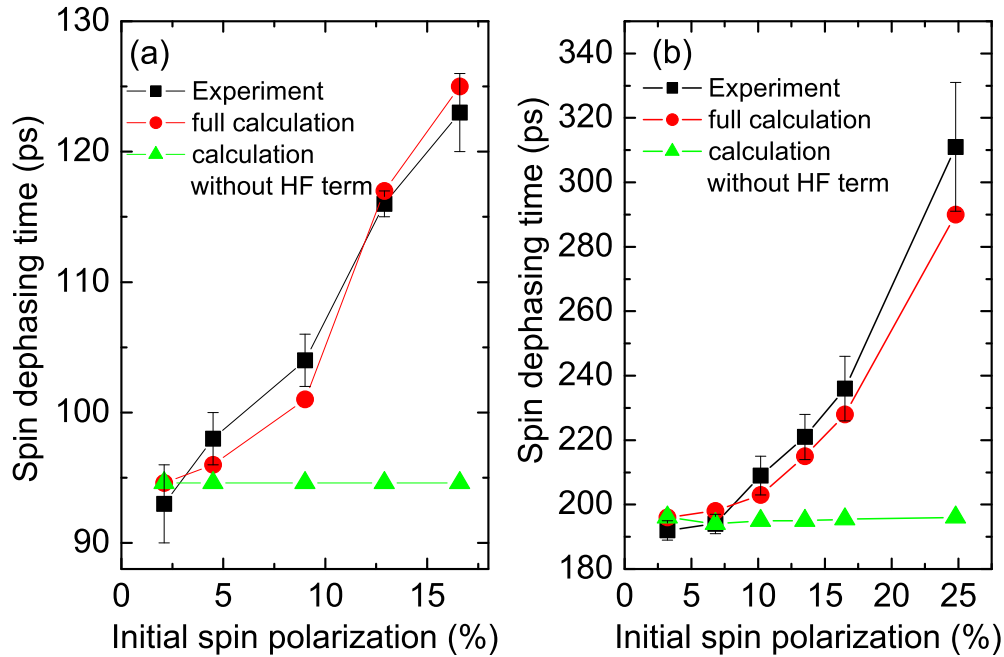,width=0.8\linewidth}}
\caption{(a) The spin dephasing times as a function of initial spin
polarization for constant, \emph{low} excitation density and
variable polarization degree of the pump beam. The measured spin
dephasing times are compared to calculations with and without the
Hartree-Fock (HF) term, showing its importance. (b) The spin dephasing times measured and calculated
for constant, \emph{high} excitation density and variable
polarization degree. From Stich et al. \cite{stich2}.}
\label{fig2}
\end{figure}

\begin{figure}[htb]
\begin{center}
\includegraphics[width=6.5cm]{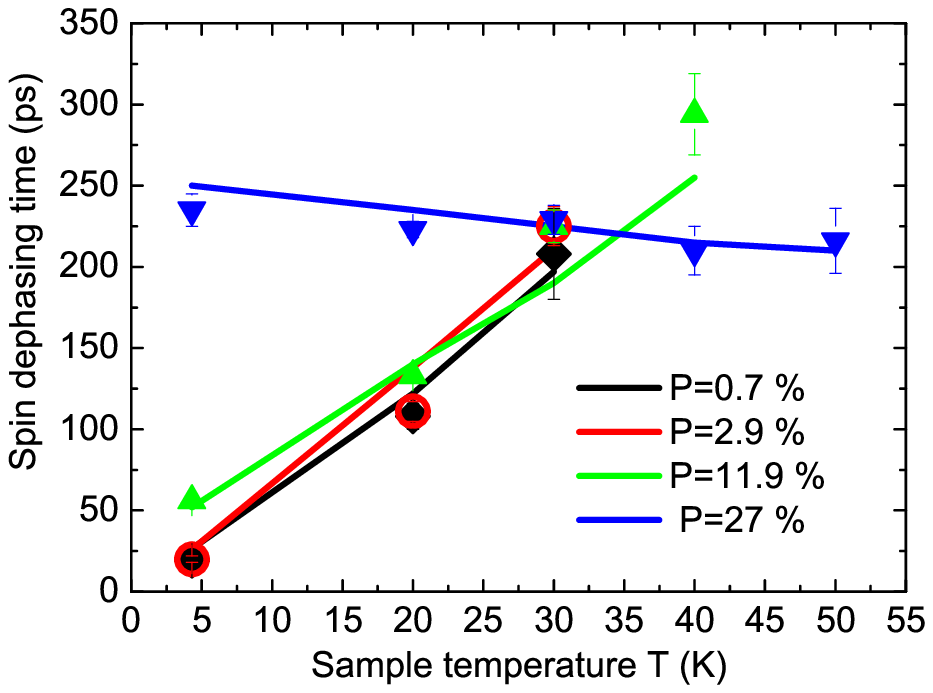}
\includegraphics[width=6.2cm]{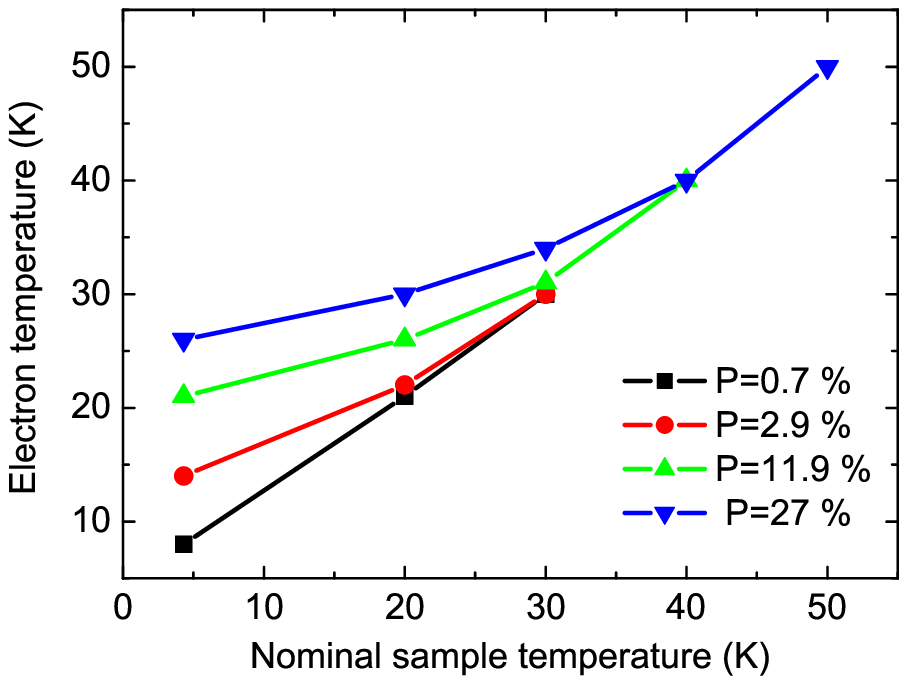}
\end{center}
\caption{(a) Spin dephasing time as a function of sample temperature, for
 different initial spin polarizations. The measured data points are represented by
 solid points, while the calculated data are represented
 by  lines of the same colour. (b) Electron temperature determined
 from intensity-dependent photoilluminance measurements as
 a function of the nominal sample temperature, for different pump beam fluence
 and initial spin polarization, under experimental conditions corresponding to
 the measurements shown in (a). The measured data points are represented by
 solid points, while the curves serve as guide to the eye. From Stich
 et al. \cite{stich2}.}
\label{fig3}
\end{figure}

\begin{figure}[htb]
\centerline{\includegraphics[width=14.cm]{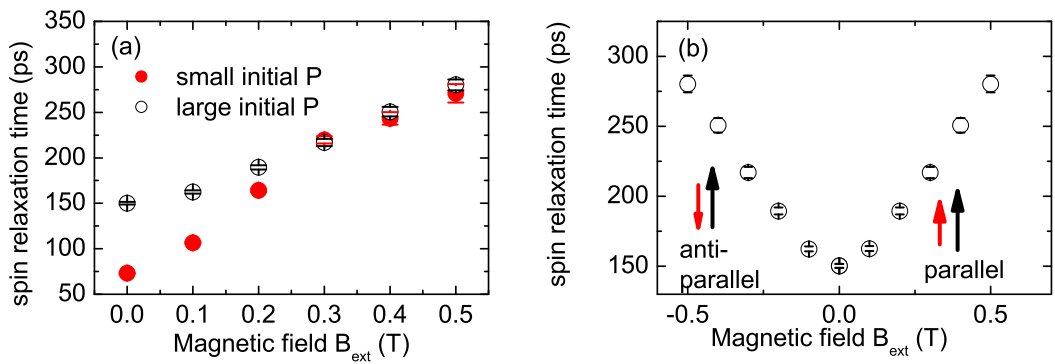}}
\caption{ (a) Spin dephasing times as a function of an
  external magnetic field perpendicular to the quantum well plane for
  small and large initial spin polarization. (b) Same as (a) for large
  initial spin polarization and both polarities of the external
  magnetic field. From Korn et al.\cite{korn}}
\label{fig4}
\end{figure}

Another important effect from the Coulomb interaction to the spin relaxation/dephasing comes
from the Coulomb Hartree-Fock contribution in the coherent terms of the kinetic spin Bloch equations.
Weng and Wu \cite{weng} first pointed out that at a high spin
polarization, the Hartree-Fock term serves as an effective magnetic field along
the $z$ axis which blocks the spin precession. As a result,
 the spin relaxation/dephasing time increases
dramatically with the spin polarization. They further
pointed out that the spin relaxation/dephasing time decreases with temperature at high spin
polarization in quantum well with small well width, which is in contrast to
the situation with small spin polarizations. These predictions have been
verified experimentally by Stich {\em et al.} in an
$n$-type (001) GaAs quantum well with high mobility \cite{stich1,stich2}.
 By changing the intensity of the circularly polarized
lasers, Stich {\em et al}. measured the spin dephasing time in a high
mobility $n$-type GaAs quantum well as a function of initial spin
polarization. Indeed they
observed an increase of the spin dephasing time with the increased
spin polarization, and the theoretical calculation based on the
kinetic spin Bloch equations nicely reproduced
the experimental findings when the Hartree-Fock term was included \cite{stich1}.
It was also shown that when the
Hartree-Fock term is removed, one does not see any increase of the
spin dephasing time. Later, they further
improved the experiment by replacing the circular-polarized laser
pumping with the elliptic polarized laser pumping. By doing
so, they were able to vary the spin polarization without changing the
carrier density. Figure.~\ref{fig2} shows the measured
spin dephasing times as function of initial spin polarization under
two fixed pumping intensities, together with the theoretical calculations with
and without the Coulomb Hartree-Fock term. Again the spin dephasing
time increases with the initial spin polarization as predicted and the
theoretical calculations with the Hartree-Fock term are in good
agreement with the experimental data \cite{stich2}. Moreover, Stich {\em et al}.
also confirmed the prediction of the temperature dependences of the
spin dephasing time at low
and high spin polarizations \cite{stich2}. Figure~\ref{fig3}(a) shows
the measured temperature dependences of the spin dephasing time at
different initial spin polarizations. As predicted, the spin
dephasing time increases with increasing temperature at small spin
polarization but decreases at large spin polarization. The theoretical
calculations also nicely reproduced the experimental data. The
hot-electron temperatures in the calculation were taken from the
experiment [Fig.~\ref{fig3}(b)].
The effective magnetic field
from the Hartree-Fock term has been measured by Zhang {\em et
al}. from the sign switch of the Kerr signal and the phase reversal
of Larmor precessions with a bias voltage in a GaAs
heterostructure \cite{zheng1}.
Korn {\em et al}. \cite{korn} also estimated the average effect by applying an
external magnetic field in the Faraday configuration, as shown in
Fig.~\ref{fig4}(a) for the same sample reported above \cite{stich1,stich2}.
They compared the spin  dephasing times of both large and
small spin polarizations as function of
external magnetic field. Due to the effective magnetic field from the
Hartree-Fock term, the spin relaxation times are different under small
external magnetic field but become identical when the magnetic field
becomes large enough. From the merging point, they estimated the mean
value of the effective magnetic field is below 0.4~T. They further
showed that this effective magnetic field from the Hartree-Fock term
cannot be compensated by the external magnetic field, because it does
not break the time-reversal symmetry and is therefore not a genuine
magnetic field, as said above. This can be seen from
Fig.~\ref{fig4}(b) that the spin relaxation time at large spin
polarization shows identical external magnetic field dependences when
the magnetic field is parallel or antiparallel to the growth
direction.

%\begin{widetext}
%\end{widetext}

We now turn to discuss the spin relaxation/dephasing far away from the equilibrium.
In fact, the spin relaxation/dephasing of high spin polarization addressed above is one
of the cases far away from the equilibrium. Another case is the spin
dynamics in the presence of a high in-plane electric field.
The spin dynamics in the presence of a high in-plane electric field
was first studied by Weng {\em et al.} \cite{weng2} in GaAs quantum well
with only the lowest subband by solving the kinetic spin Bloch equations.
To avoid the ``runaway'' effect \cite{run},
 the electric field was calculated upto 1~kV/cm. Then Weng and Wu further
introduced the second subband into the model and the in-plane electric
field was increased upto 3~kV/cm \cite{weng6}. Zhang {\em et al.}
included $L$ valley and the electric field was further increased upto
7~kV/cm  \cite{zhang1}.
The effect of in-plane electric field to the spin relaxation in system
with strain was investigated by Jiang and Wu \cite{ljiang}. Zhou
{\em et al}. also investigated the electric-field effect at low
lattice temperatures \cite{zhou1}.

The in-plane electric field leads to two effects: i) It shifts the
center-of-mass of electrons to ${\bf k}_d=m^\ast{\bf v}_d=m^\ast\mu{\bf E}$ with
$\mu$ representing the mobility, which further induces an effective
magnetic field via the D'yakonov-Perel' term \cite{weng2}.
ii) The in-plane electric field also leads to the hot-electron
effect \cite{conwell}. The first effect induces
a spin precession even in the absence of any external magnetic
field and the spin precession frequency changes with the direction of
the electric field in the presence of an external magnetic
field \cite{weng2,zhang1}. The second effect enhances both the inhomogeneous
broadening and the scattering, two competing effects leading to
rich electric-field dependence of the spin relaxation/dephasing and thus spin
manipulation \cite{weng2,weng6,zhang1,ljiang,zhou1,jiang2,jiang4}.

Finally we address some issues of which the kinetic-spin-Bloch-equation approach
gives qualitatively different predictions from those widely used in the
literature. These issues include the Bir-Aronov-Pikus mechanism, the Elliot-Yafet mechanism and
some density/temperature dependences of the spin relaxation/dephasing time.

\begin{figure}[ht]
\centerline{\includegraphics[width=7cm]{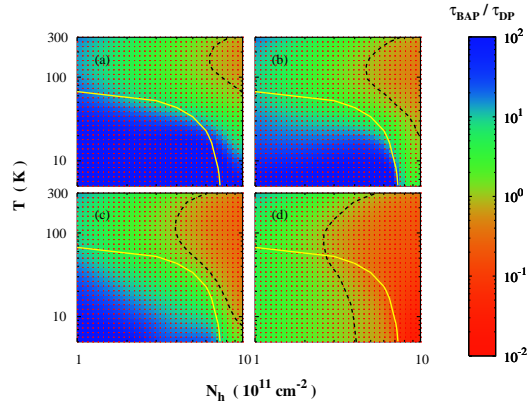}}
  \caption{ Ratio of the spin relaxation time due to the Bir-Aronov-Pikus mechanism
     to that due to the D'yakonov-Perel' mechanism, $\tau_{\rm BAP}/\tau_{\rm DP}$,
     as function of temperature and hole density with
     (a) $N_i=0$, $N_{ex}=10^{11}$~cm$^{-2}$;
     (b) $N_i=0$, $N_{ex}=10^{9}$~cm$^{-2}$; (c) $N_i=N_h$,
     $N_{ex}=10^{11}$~cm$^{-2}$; (d) $N_i=N_h$, $N_{ex}=10^{9}$~cm$^{-2}$.
     The black dashed curves indicate the cases satisfying
     $\tau_{\rm BAP}/\tau_{\rm DP}=1$. Note the smaller the
     ratio $\tau_{\rm BAP}/\tau_{\rm DP}$ is, the more
     important the Bir-Aronov-Pikus mechanism becomes.
     The yellow solid curves indicate the cases satisfying
     $\partial_{\mu_h}[N_{{\rm LH}^{(1)}}+N_{{\rm HH}^{(2)}}]/\partial_{\mu_h}N_h=0.1$.
     In the regime above the yellow curve the multi-hole-subband
     effect becomes significant. From Zhou {\em et al}. \cite{yzhou}.}
\label{fig5}
\end{figure}

It has long been believed in the literature that for electron relaxation/dephasing,
 the Bir-Aronov-Pikus mechanism is
dominant at low temperature in $p$-type samples and has important
contribution to intrinsic sample with high
photo-excitation \cite{bap1,bap2,bap3,bap4,bap5,bap6,bap7}. These conclusion
was made based on the single-particle Fermi golden rule. Zhou and We reexamined
the problem using the kinetic-spin-Bloch-equation
 approach \cite{zhou2}. They pointed out that the
Pauli blocking was overlooked in the Fermi Golden rule approach.
When electrons are in the non-degenerate limit, the results calculated from the
 Fermi Golden rule approach are valid. However, at low temperature, electrons
can be degenerate  and the Pauli blocking becomes very important.
As a result, the previous approaches always overestimated the importance of the
Bir-Aronov-Pikus mechanism at low temperature. Moreover, the previous single-particle theories
underestimated the contribution of the D'yakonov-Perel' mechanism by neglecting the Coulomb
scattering. Both made the Bir-Aronov-Pikus mechanism dominate the spin relaxation/dephasing at low temperature.
Later, Zhou {\em et al}. performed a thorough investigation of
electron spin relaxation in $p$-type (001) GaAs quantum wells by
varying impurity, hole and photo-excited electron
densities over a wide range of values \cite{yzhou}, under the idea that very
high impurity density and very low photo-excited electron density may
effectively suppress the importance of the D'yakonov-Perel' mechanism and the Pauli
blocking. Then the relative importance of the Bir-Aronov-Pikus and D'yakonov-Perel'
mechanisms may be reversed. This indeed happens as shown in the
phase-diagram-like picture in Fig.~\ref{fig5} where the relative
importance of the Bir-Aronov-Pikus and D'yakonov-Perel' mechanisms is plotted as function of hole
density and temperature at low and high impurity densities and
photo-excitation densities. For the situation of high hole density
they even included multi-hole subbands as well as the light hole band. It
is interesting to see from the figures that at relatively high
photo-excitations, the Bir-Aronov-Pikus mechanism becomes more important than the
D'yakonov-Perel' mechanism only at high hole densities and high temperatures (around
hole Fermi temperature) when the impurity is very low [zero in
Fig.~\ref{fig5}(a)]. Impurities can suppress the D'yakonov-Perel' mechanism and
hence enhance the relative importance of the Bir-Aronov-Pikus mechanism. As a
result, the temperature regime is extended, ranging from  the hole
Fermi temperature to the electron Fermi temperature for high hole
density. When the photo-excitation is weak so that the Pauli blocking
is less important, the temperature regime where the Bir-Aronov-Pikus mechanism is
important becomes wider compared to the high excitation case. In
particular, if the impurity density is high enough and the
photo-excitation is so low that the electron Fermi temperature is
below the lowest temperature of the investigation, the Bir-Aronov-Pikus mechanism
can dominate the whole temperature regime of the investigation at
sufficiently high hole
density, as shown in Fig.~\ref{fig5}(d). The corresponding spin
relaxation times of each mechanism under high or low impurity and
photo-excitation densities are demonstrated in
Fig.~\ref{fig5}. They also discussed the density dependences of
spin relaxation with some intriguing properties related to the high
hole subbands \cite{yzhou}.
The predicted Pauli-blocking effect in the Bir-Aronov-Pikus mechanism has been
partially demonstrated experimentally by Yang {\em et al}. \cite{yang}
They showed by increasing the pumping
density, the temperature dependence of the spin dephasing time
deviates from the one from the Bir-Aronov-Pikus mechanism and the peaks at high
excitations agree well with those predicted by Zhou and Wu \cite{zhou2}.

Another widely accepted but incorrect conclusion is related to the Elliot-Yafet mechanism.
It is widely accepted in the literature that the Elliot-Yafet mechanism
dominates spin relaxation in $n$-type bulk III-V semiconductor
at low temperature, while the D'yakonov-Perel'
mechanism is important at high temperature \cite{song,Meier:1984, Zutic2004:RMP, Fabian2007:APS, rev}.
Jiang and Wu pointed out that the previous understanding are based on
the formula that can only be used in the nondegenerate limit. Moreover, the
momentum relaxation rates are calculated via the approximated formula for
mobility \cite{song}. By performing an accurate calculation via the
kinetic-spin-Bloch-equation
 approach, they showed that the Elliot-Yafet mechanism is {\em not} important in
III-V semiconductors, including even the narrow-band InAs
and InSb \cite{jiang4}.
Therefore, the D'yakonov-Perel' mechanism is the only spin relaxation mechanism for $n$-type
III-V semiconductors in metallic regime.

\begin{figure}
\centerline{\includegraphics[width=7.cm]{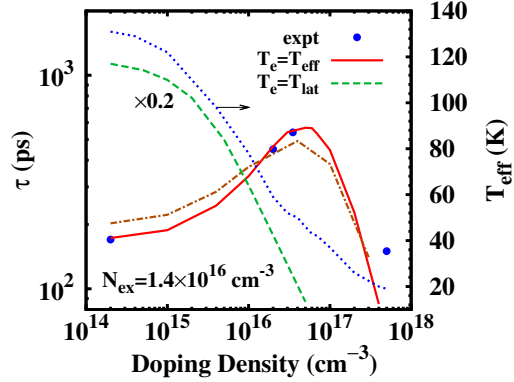}}
\caption{Electron spin
  relaxation times from the calculation via the kinetic-spin-Bloch-equation
 approach (red
  solid curve) and from the experiment\cite{kruss} (blue $\bullet$) as
  function of the doping density. The green dashed curve shows the
  results without hot-electron effect. The
  hot-electron temperature used in the computation is plotted as the blue
  dotted curve (Note the scale is on the right hand side of the
frame). The chain curve is the calculated spin relaxation time with a fixed
hot-electron  temperature  80~K, and $N_{ex}=6\times10^{15}$~cm$^{-3}$. From
Shen \cite{shen}.}
\label{fig6}
\end{figure}

Jiang and We have further predicted a peak in the density dependence
of the spin relaxation/dephasing time in $n$-type III-V semiconductors where the spin relaxation/dephasing
is limited by the D'yakonov-Perel' mechanism \cite{jiang4}.
Previously, the nonmonotonic density
dependence of spin lifetime was observed in low-temperature
($T\lesssim$5~K) experiments, where the localized electrons play a
crucial role and the electron system is in the insulating regime or
around the metal-insulator transition point \cite{dzhi}.
Jiang and Wu found, for
the first time, that the spin lifetime in {\em metallic} regime is also
{\em nonmonotonic}. Moreover, they pointed out that it is a {\em
 universal} behavior for {\em all} bulk III-V semiconductors at {\em
 all} temperature where the peak is located at $T_F\sim T$ with $T_F$
being the electron Fermi temperature.  The
underlying physics for the nonmonotonic density dependence in metallic
regime can be understood as following: In the nondegenerate regime,
as the distribution is the Boltzmann one,  the density
dependence of the inhomogeneous broadening is marginal. However, the
scattering increases with the density. Consequently the spin relaxation/dephasing time
increases with the density. However, in the degenerate regime, due to the Fermi
distribution, the inhomogeneous broadening increases with the density
much faster than the scattering does. As a result, the spin relaxation/dephasing time decreases
with the density.  Similar behavior was also found in
two-dimensional system \cite{jiang1,zhang3}, where the
underlying physics is similar. The predicted peak was later observed
by Krau\ss\ {\em et al}. \cite{kruss} as shown
in Fig.~\ref{fig6} where theoretical calculation based on the kinetic spin Bloch equations
nicely reproduced the experimental data by Shen \cite{shen}.

\subsection{Spin diffusion/transport}

By solving the kinetic spin Bloch equations together with the Poisson equation
self-consistently, one is able to obtain all the transport properties
such as the mobility, charge diffusion length and spin
diffusion/injection length without any fitting parameter. It was first
pointed out by Weng and Wu \cite{weng3} that the drift-diffusion equation
approach is inadequate in accounting for the spin
diffusion/transport. It is important to include the off-diagonal term between
opposite spin bands $\rho_{{\mathbf k} \uparrow \downarrow}$
in studying the spin diffusion/transport. With this term, electron
spin precesses along the diffusion and therefore ${\mathbf k} \cdot
{\mathbf \nabla}_{{\mathbf r}} \rho_{{\mathbf k}} ({\mathbf r}, t)$  in the
diffusion terms $\dot \rho_{\mathbf k}(\mathbf{r}, t)|_{\mathtt{dif}}$
provides an additional inhomogeneous broadening. With this
additional inhomogeneous broadening, any scattering, including the
Coulomb scattering, can cause an irreversible spin relaxation/dephasing
\cite{weng3}. Unlike
the spin precession in the time domain where the inhomogeneous
broadening is determined by the effective magnetic field
from the D'yakonov-Perel' term, ${\mathbf h} ({\mathbf k})$, in spin diffusion
and transport it is determined by
\begin{equation}
\bgreek{\Omega}_{\bf k}=|g \mu_{B} {\mathbf B}+{\mathbf h}({\mathbf k})|/ k_{x},
\label{ome}
\end{equation}
provided the diffusion is along the $x$-axis \cite{cheng}. Here the magnetic
field is in the Voigt configuration.
Therefore, even in the absence of the
D'yakonov-Perel' term ${\bf h}({\bf k})$, the magnetic field {\em alone} can
provide an inhomogeneous broadening and leads to
the spin relaxation/dephasing in spin diffusion and transport. This was first pointed out by
Weng and Wu back to 2002 \cite{weng3} and   has been
realized experimentally by Appelbaum \rm {et al.}
 in bulk silicon \cite{app1,app2}, where there is no D'yakonov-Perel' spin-orbit coupling
due to the center inversion symmetry. Zhang  and Wu further investigated
the spin diffusion and transport in symmetric Si/SiGe quantum
wells \cite{zhang4}.

When $B=0$ but the D'yakonov-Perel' term is present, then the inhomogeneous broadening
for spin diffusion and transport is determined by
$\bgreek{\Omega}_{\bf k}={\mathbf h}({\mathbf k})/ k_{x}$. In (001) GaAs quantum well
where the D'yakonov-Perel' term is determined by the Dresselhaus term
\cite{dress},
the average of $\bgreek{\Omega}_{\bf k}$ reads
$\langle \bgreek{\Omega}_{\mathbf{k}}\rangle=C(\langle k_y^2\rangle -\langle
k_z^2\rangle,0,0)$ with $C$ being a constant. For electrons
in quantum well, this value is not
zero. Therefore, the spacial spin oscillation due to the Dresselhaus
effective magnetic field survives even at high temperature when the
scattering is strong. This effect was  first predicted by
Weng and Wu by showing  a spin pulse
can oscillate along the diffusion in the absence of the magnetic field
at very high temperature \cite{weng4}. Detailed studies were
carried out later on this effect \cite{weng5,ljiang1,cheng}.
The spin oscillation without any applied magnetic field in the
transient spin transport was later
observed experimentally by Crooker and Smith in
strained bulk system \cite{crooker}.
differing from the two-dimensional case, in bulk the average of
$\bgreek{\Omega}_{\mathbf{k}}$ from the Dresselhaus term is zero, since
$\langle\bgreek{\Omega}_{\mathbf{k}}\rangle=C(\langle
k_y^2\rangle-\langle k_z^2\rangle,0,0)=0$ due to the symmetry in
the $y$- and $z$-directions. This is consistent
with the experimental
result that there is no spin oscillation for the system without
stress. However, when the stress is applied, an additional spin-orbit
coupling,  namely the coupling of electron spins to the strain tensor,
appears, which is linear in momentum \cite{Meier:1984}.
This additional spin-orbit coupling
also acts as an effective magnetic field.
Therefore, once the stress is
applied, one can observe spacial spin oscillation
 even when there is no applied magnetic field \cite{crooker}.

Cheng and Wu further developed a new numerical scheme to calculate the spin
diffusion/transport in GaAs quantum wells
with very high accuracy and speed \cite{cheng}. It was
discovered that due to the scattering, especially the Coulomb
scattering, $T_{2}=T_{2}^{\ast}$ is valid even in the spacial domain.
This prediction remains yet to be verified experimentally.
Moreover, as the inhomogeneous broadening in spin diffusion is determined by
$|{\mathbf h} ({\mathbf k})|/k_{x}$ in the
absence of magnetic field, the period of the spin oscillations along
the $x$-axis is independent on the electric field perpendicular
to the growth direction of the quantum well \cite{cheng}, which is
different from the spin precession rate in the time domain \cite{weng2}. This
is consistent with the experimental findings by
Beck {\em et al}. \cite{beck}.

\begin{figure}
\centerline{\includegraphics[width=6.cm]{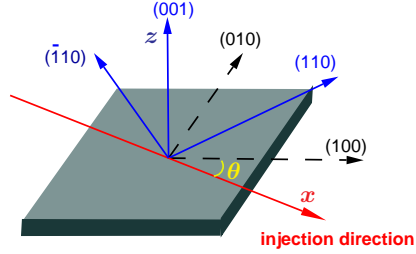}}
\caption{Schematic of the different directions considered
for the spin polarizations [(110), ($\bar{1}10$) and (001)-axes]
and spin diffusion/injection
($x$-axis). From Cheng {\em et al.} \cite{cheng1}.}
\label{fig7}
\end{figure}

\begin{figure}[hbt]
\centerline{\includegraphics[width=6cm]{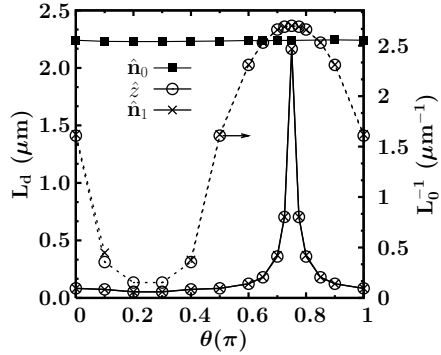}}
\caption{Spin diffusion length $L_d$ (solid curves) and the inverse of
the spin oscillation period $L_0^{-1}$ (dashed curves)
for identical Dresselhaus and Rashba coupling strengths
 as functions of the injection
direction for different spin polarization directions
$\hat{\mathbf n}_0$, $\hat{\mathbf z}$ and $\hat{\mathbf n}_1$
($\hat{\mathbf n}_1=\hat{\mathbf z}\times\hat{\mathbf n}_0$, i.e., crystal
direction $[\bar{1}10]$) at
$T=200$\ K. It is noted that the scale of the spin oscillation
period is on the right hand side of the frame.
From Cheng  {\em et al}. \cite{cheng1}.}
\label{fig8}
\end{figure}

Cheng {\em et al.} applied the kinetic-spin-Bloch-equation
 approach to study the spin transport in
the presence of competing Dresselhaus and Rashba fields \cite{cheng1}.
When the Dresselhaus and Bychkov-Rashba
\cite{Bychkov1984:JETPL} terms are both important in
semiconductor quantum well, the total effective magnetic field can be
highly anisotropic and spin dynamics is also highly anisotropic in
regards to the spin polarization \cite{golub}.
For some special polarization direction, the spin relaxation time
is extremely large \cite{golub,golub1,loss,winkler}.
For example, if the coefficients of the
linear Dresselhaus and Bychkov-Rashba terms are equal
to each other in (001) quantum well of small well width
and the cubic Dresselhaus term is not important,
the effective magnetic field is along the [110] direction for all
electrons. For the spin components perpendicular to the [110] direction, this
effective magnetic field flips the spin and leads to a finite spin
relaxation/dephasing time. For spin along the [110] direction, this effective
magnetic field can not flip it. Therefore, when the spin
polarization is along the [110] direction,
the Dresselhaus and Bychkov-Rashba terms can not cause any spin relaxation/dephasing.
When the cubic Dresselhaus term is taken into
account, the spin dephasing time for spin polarization along the [110]
direction is finite but still much larger than other directions \cite{cheng2}.
The anisotropy in the spin direction is also expected in spin diffusion and
transport.
When the Dresselhaus and Bychkov-Rashba terms are comparable, the spin
injection length $L_d$ for the spin polarization
perpendicular to [110] direction is usually much shorter than
that for the spin polarization along [110] direction.
In the ideal case when there are only the linear Dresselhaus
and Bychkov-Rashba terms with identical strengths,
spin injection length for spin polarization parallel to the
[110] direction becomes infinity \cite{golub1,loss}.
This effect has promoted
Schliemann {\em et al.} to propose the nonballistic spin-field-effect
transistor \cite{loss}. In such a transistor, a gate
voltage is used to tune the strength of the Bychkov-Rashba term and therefore
control the spin injection length.
However, Cheng {\em et al.} pointed out that  spin diffusion and
transport actually involve both the spin polarization
and spin transport directions \cite{cheng1}. The latter has long been
overlooked in the literature. In the
kinetic-spin-Bloch-equation approach, this direction corresponds to the
spacial gradient in the diffusion term [$\dot \rho_{\mathbf k}(\mathbf{r},
 t)|_{\mathtt{dif}}$] and the electric field in the drifting term
[$\dot \rho_{\mathbf k}(\mathbf{r}, t)|_{\mathtt{dr}}$]. The
importance of the spin transport direction has not been realized until
Cheng {\em et al.} pointed out that
the spin transport is highly anisotropic not only in the sense of
the spin polarization direction but also in the spin transport direction
when the Dresselhaus and Bychkov-Rashba effective magnetic fields
are comparable \cite{cheng1}.
They even predicted that in (001) GaAs quantum well with identical
linear Dresselhaus and Bychkov-Rashba coupling strengths,
the spin injection along $[\bar{1}10]$ or
$[110]$ \cite{footnote} can be infinite {\em regardless of} the
direction of the spin polarization. This can be easily seen from
the inhomogeneous broadening Eq.~(\ref{ome}) which well defines the spin
diffusion/transport properties. For the spin diffusion/transport in a (001) GaAs
quantum well with identical Dresselhaus and Bychkov-Rashba strengths (the schematic is
shown in Fig.~\ref{fig7} with the transport direction chosen along the
$x$-axis), the inhomogeneous broadening is given by \cite{cheng1}
\begin{eqnarray}
\bgreek{\Omega}_{\mathbf{k}}& =&
\biggl\{2\beta
\left(
  \sin(\theta-\frac{\pi}{4})+\cos(\theta-\frac{\pi}{4})\frac{k_y}{k_x}\right)
\hat{\mathbf n}_0\nonumber\\
&&\mbox{}\hspace{-0.6cm}  +
\gamma(\frac{k_x^2-k_y^2}{2}\sin
2\theta+k_xk_y\cos 2\theta)
\bigl(\frac{k_y}{k_x},\;
-1, \;
0\bigr)\biggr\},
\label{eq:trans:omega_ab}
\end{eqnarray}
with $\theta$ being the angle between the spin transport direction ($x$-axis)
and [001] crystal direction.
It can be splitted into
two parts: the zeroth-order
term (on $k$) which is always along the same direction of
$\hat{\mathbf  n}_0$ and the second-order term which comes from the
cubic Dresselhaus term.
If the cubic Dresselhaus term is omitted, the effective magnetic fields
for all $\mathbf{k}$ states align along $\hat{\mathbf{n}}_0$ (crystal
[110]) direction. Therefore, if the spin polarization is along
$\hat{\mathbf{n}}_0$, there is no spin relaxation even in
the presence of scattering since there is no spin precession.
Nevertheless, it is interesting to see from
Eq.~(\ref{eq:trans:omega_ab}) that when
$\theta=3\pi/4$, i.e., the spin transport is along the [$\bar{1}10$] direction,
$\bgreek{\Omega}_{\mathbf{k}}=2m^{\ast}\beta\hat{\mathbf{n}}_0$ is independent
on $\mathbf{k}$ if the cubic Dresselhaus term is neglected.
Therefore, in this special spin transport direction, there is no
inhomogeneous broadening in the spin transport for {\em any} spin
polarization.
The spin injection length is therefore infinite
regardless of the direction of spin polarization.
This result is highly counterintuitive, considering that the
spin relaxation times for the spin components perpendicular to the
effective magnetic field are
finite in the spacial uniform system.
The surprisingly contradictory results,
{i.e.}, the finite spin relaxation/dephasing time versus the infinite spin
injection length, are due to the difference in the inhomogeneous
broadening in spacial uniform and non-uniform systems.
For genuine situation, due to the presence of the cubic term,
the spin injection length is still finite and the maximum
spin injection length does not happen at the identical Dresselhaus
and Bychkov-Rashba coupling strengths, but shifted by a small amount due to the
cubic term \cite{cheng1}. However, there is strong
anisotropy in regards to the spin polarization and spin injection
direction, as shown in Fig.~\ref{fig8}.
This predication has not yet been realized experimentally. However,
very recent experimental findings on spin helix
\cite{Bernevig2006:PRL,Koralek2009:N} have
provide strong evidence to support this predication \cite{shen1}.

\begin{figure}[htpb]
\centerline{ \includegraphics[width=5cm]{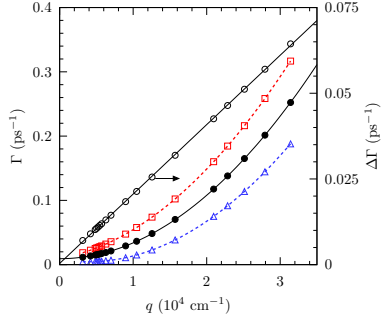}}
 \caption{$\Gamma=(\Gamma_++\Gamma_-)/2$ and
     $\Delta\Gamma=(\Gamma_+-\Gamma_-)/2$
     {\em vs}. $q$ at  $T=295$\ K.
     Open boxes/triangles are the relaxation rates $\Gamma_{+/-}$
     calculated from the full kinetic spin Bloch equations.
     Filled/open circles represent $\Gamma$ and $\Delta\Gamma$ respectively.
     Noted that the scale for $\Delta\Gamma$ is on the
     right hand side of the frame. The solid curves are the fitting to
     $\Gamma$ and $\Delta\Gamma$ respectively. The dashed curves are
     guide to eyes.  From Weng {\em et al}. \cite{weng7}.
   }
   \label{fig9}
 \end{figure}

Now we turn to the problem of spin grating.
Transient spin grating, whose spin polarization varies periodically in
real space, is excited optically by two non-collinear coherent light
beams with orthogonal linear polarization
\cite{grat1,grat2,grat3,Koralek2009:N}.
Transient spin grating technique
can be used to study the spin transport since it can directly probe
the decay rate of nonuniform spin distributions.
Spin diffusion coefficient $D_s$ can be obtained from the transient
spin grating experiments \cite{grat1,grat2,grat3}.
In the literature, the drift-diffusion model was employed to extract
$D_s$ from the experimental data.
With the drift-diffusion model,
the transient spin grating was predicted to decay exponentially
with time with a decay rate of
$\Gamma_q=D_sq^2+1/\tau_s$, where $q$ is the wavevector of the spin
grating and $\tau_s$ is the spin relaxation time \cite{grat1,grat2}.
However, this result is not accurate since it neglects the spin
precession which plays an important role in spin
transport as first pointed out by Weng and Wu \cite{weng2}.
Indeed, experimental results show that the decay of transient
spin grating takes a double-exponential form instead of single
exponential one \cite{grat1,grat3,Koralek2009:N}. Also the relation which
relates the  spin injection length with the spin diffusion coefficient $D_s$
and the spin relaxation time $\tau_s$ $L_s=2\sqrt{D_s\tau_s}$  from the
drift-diffusion model should be checked. In fact, if this relaxation is correct,
the above  prediction of infinite spin injection length at certain
spin injection direction for any spin polarization in the presence of
identical Dresselhaus and Bychkov-Rashba coupling strengths cannot be correct.

\begin{figure}[htpb]
\centerline{  \includegraphics[width=7cm]{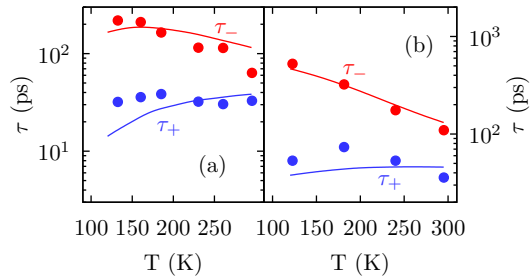}}
   \caption{ Spin relaxation times $\tau_\pm$
     vs. temperature
     for (a) high-mobility sample with $q=0.58\times 10^4$\ cm$^{-1}$
     and (b) low-mobility sample with $q=0.69\times
     10^4$\ cm$^{-1}$. The dots are the experiment data from
     Ref.~\cite{grat3}.
     From Weng {\em et al}. \cite{weng7}.
 }
   \label{fig10}
 \end{figure}

Weng {\em et al.} studied this problem from the kinetic-spin-Bloch-equation
 approach \cite{weng7}.
By first solving the kinetic spin Bloch equations analytically by including only the elastic
scattering, {\em i.e.}, the electron-impurity scattering, they
showed that the transient spin grating should decay double
exponentially with two decay rates $\Gamma_\pm$. In fact, non of the
rates is quadratic in $q$. However, the average of them reads \cite{weng7}
\begin{equation}
\Gamma=(\Gamma_++\Gamma_-)/2=Dq^2+1/\tau^\prime_s
%(1/\tau_s+1/\tau_{{s1}})/2,
\label{Gamma}
\end{equation}
with
$1/\tau^\prime_s=(1/\tau_s+1/\tau_{{s1}})/2$,
which differs from
the current widely used formula by replacing the spin decay rate by the
average of the in- and out-of-plane relaxation rates.
The difference of these two decay rates is a linear function of the
wavevector $q$ when $q$ is relatively large:
\begin{equation}
  \label{DGamma}
  \Delta\Gamma=cq+d,
\end{equation}
with $c$ and $d$ being two constants.
The steady-state spin injection length $L_s$ and spin precession period $L_0$
are then \cite{weng7}
\begin{eqnarray}
\label{ls}
L_s&=&2D_s/\sqrt{|c^2-4D_s(1/\tau^{\prime}_s-d)|},\\
L_0&=&2D_s/c.
\label{l0}
\end{eqnarray}
They further showed that the above relations Eqs.~(\ref{Gamma}) and
(\ref{DGamma})
are valid even including the inelastic electron-electron and electron phonon
scatterings by solving the full kinetic spin Bloch equations, as shown in
Fig.~\ref{fig9}. A good
 agreement with the experimental data \cite{grat3}
 of double exponential decays $\tau_\pm=\Gamma_\pm^{-1}$
are shown in Fig.~\ref{fig10}. Finally it was shown that
the infinite spin injection length predicted by Cheng {\em et al.}
in the presence of identical Dresselhaus and Bychkov-Rashba coupling
strengths \cite{cheng1} addressed above
can exactly be obtained from Eq.~(\ref{ls}) as in that special
case $\tau_s^\prime=\tau_s$, $d=0$ and $c=2\sqrt{D_s/\tau_s}$ \cite{weng7}.
However, $\sqrt{D_s\tau_s}$ always remains finite unless $\tau_s=\infty$.
Therefore, Eqs.~(\ref{Gamma}-\ref{l0}) give the correct way to extract the
spin injection length from the spin grating measurement.

\noindent {\bf Acknowledgements}

This work was supported by DFG SFB 689 and SPP1286, 
Natural Science Foundation of China under Grant
No.~10725417, the National Basic Research Program of China under Grant
No.~2006CB922005 and the Knowledge Innovation Project of Chinese Academy of
Sciences.

\bibliographystyle{bibit1}

%\bibliography{../../references_master}

\begin{thebibliography}{0}

\bibitem{Elliott1954:PR} R. J. Elliott, ``Theory of the effect of spin-orbit coupling on magnetic resonance in
            some semiconductors'', Phys. Rev. {\bf 96}, 266 (1954).

\bibitem{Yafet:1963} Y. Yafet in {\it Solid State Physics}, Vol 14, edited by F. Seitz and D. Turnbull
(Academic, New York, 1963), p. 2.

\bibitem{Dyakonov1972:SPSS} M. I. D'yakonov and V. I. Perel',
``Spin relaxation of conduction electrons in noncentrosymmetric
                 semiconductors'' Sov. Phys. Solid State {\bf 13}, 3023 (1971).

\bibitem{Bir1976:SPJETP} G. L. Bir, A. G. Aronov, and G. E. Pikus, ``Spin relaxation of electrons due to scattering by holes'',
Sov. Phys. JETP {\bf 42}
705 (1976).

\bibitem{Meier:1984} F. Meier and B. P. Zakharchenya (Eds.), {\it Optical Orientation}
(North-Holland, New York, 1984).

\bibitem{Fabian1999:JVST} J. Fabian and S. Das Sarma, ``Spin relaxation of conduction electrons'',
J. Vac. Sci. Technol. B {\bf 17},
1708 (1999).

\bibitem{Zutic2004:RMP} I. \v{Z}uti\v{c}, J. Fabian, and S. Das Sarma, ``Spintronics: Fundamentals
and Applications'', Rev. Mod. Phys. {\bf 76},
323 (2004).

\bibitem{Fabian2007:APS} J. Fabian, A. Matos-Abiague, C. Ertler, P. Stano, and I. \v{Z}uti\v{c},
``Semiconductor spintronics'', Acta Phys. Slov. {\bf 57}, 565 (2007).

\bibitem{wu-review} M. W. Wu, J. H. Jiang,
and M. Q. Weng, ``Spin dynamics in semiconductors'', Phys. Rep. {\bf 493}, 61
(2010).

\bibitem{Bychkov1984:JETPL} Y. A. Bychkov and E. I. Rashba, ``Properties of a {2D} electron gas with lifted spectral degeneracy'',
JETP Lett. {\bf 39}, 78 (1984).

\bibitem{Dresselhaus1955:PR} G. Dresselhaus, ``Spin-orbit coupling effects in zinc blende structures'',
Phys. Rev. 100, 580 (1955).

\bibitem{Dyakonov1986:SPS} M. I. Dyakonov and V. Y. Kachorovskii, ``Spin relaxation of two-dimensional electrons
        in noncentrosymmetric semiconductors'',  Sov. Phys. Semicond. {\bf 20},
110 (1986).

\bibitem{Bernevig2006:PRL} B. A. Bernevig, J. Orenstein, and S. C. Zhang, ``Exact SU(2) symmetry and persistent spin helix in
       a spin-orbit coupled system'',  Phys. Rev. Lett. \textbf{97}, 236601 (2007)..

\bibitem{Koralek2009:N} J. D. Koralek, C. P. Weber, J. Orenstein, B. A. Bernevig, S. C. Zhang, S. Mack,
and D. D. Awschalom, ``Emergence of the persistent spin helix in semiconductor quantum wells '', Nature \textbf{458}, 610 (2009).

\bibitem{Cheng2010:PRL} J. L. Cheng, M. W. Wu, and J. Fabian, ``The spin relaxation of conduction electrons in silicon'',
Phys. Rev. Lett. \textbf{104}, 016601 (2010).

\bibitem{Lepine1970:PRB} D. J. Lepine, ``Spin resonance of localized and delocalized electrons in
                phosphorus-doped silicon between 20 and 30 {K}'', Phys. Rev. B {\bf 2}, 2429 (1970).

\bibitem{Huang2007:PRL} B. Huang, and I. Appelbaum, ``Coherent Spin Transport through a 350 Micron Thick Silicon Wafer'',
Phys. Rev. Lett. \textbf{99}, 177209 (2007).



\bibitem{wu}M. W. Wu and H. Metiu, ``Kinetics of spin coherence of
  electrons in an undoped semiconductor quantum well'', Phys. Rev. B {\bf 61}, 2945 (2000).
\bibitem{wu2}M. W. Wu and C. Z. Ning, ``A novel mechanism for spin
  dephasing due to spin-conserving scatterings'', Eur. Phys. J. B {\bf 18}, 373
  (2000).
\bibitem{wu3}M. W. Wu, ``Spin Dephasing Induced by Inhomogeneous
  Broadening in D'yakonov-Perel' Effect in a $n$-doped GaAs Quantum
  Well'', J. Phys. Soc. Jpn. {\bf 70}, 2195 (2001).
\bibitem{weng}M. Q. Weng and M. W. Wu, ``Spin dephasing in $n$-type
  GaAs quantum wells'', Phys. Rev. B {\bf 68}, 075312 (2003).
\bibitem{weng2}M. Q. Weng, M. W. Wu, and L. Jiang, ``Hot-electron
  effect in spin dephasing in  $n$-type GaAs quantum wells'', Phys. Rev. B {\bf 69},
  245320 (2004).
\bibitem{weng3}M. Q. Weng and M. W. Wu, ``Longitudinal spin
  decoherence in spin diffusion in semiconductors'', Phys. Rev. B {\bf 66}, 235109
  (2002).
\bibitem{weng4}M. Q. Weng and M. W. Wu, ``Kinetic theory of spin
  transport in $n$-type semiconductor quantum wells'', J. Appl. Phys. {\bf 93}, 410
  (2003).
\bibitem{weng5}M. Q. Weng, M. W. Wu, and Q. W. Shi, ``Spin
  oscillations in transient diffusion of a spin pulse in $n$-type
  semiconductor quantum wells'', Phys. Rev. B {\bf 69},
  125310 (2004).
\bibitem{ya}E. Ya. Sherman, ``Random spin--orbit coupling and spin
  relaxation in symmetric quantum wells'', Appl. Phys. Lett. {\bf 82}, 209 (2003).
\bibitem{haug}H. Haug and A. P. Jauho, {\it Quantum Kinetics in Transport
  and Optics of Semiconductors} (Springer, Berlin, 1996).
\bibitem{weng6}M. Q. Weng and M. W. Wu, ``Multisubband effect in spin
  dephasing in semiconductor quantum wells'', Phys. Rev. B {\bf 70}, 195318 (2004).
\bibitem{zhou1}J. Zhou, J. L. Cheng, and M. W. Wu, ``Spin relaxation
  in $n$-type GaAs quantum wells from a fully microscopic approach'', Phys. Rev. B {\bf 75}, 045305 (2007).
\bibitem{zhou2}J. Zhou and M. W. Wu, ``Spin relaxation due to the
  Bir-Aronov-Pikus mechanism in intrinsic and $p$-type GaAs quantum
  wells from a fully microscopic approach'', Phys. Rev. B {\bf 77}, 075318 (2008).
\bibitem{jiang1}J. H. Jiang, Y. Zhou, T. Korn, C. Sch\"uller, and
  M. W. Wu, ``Electron spin relaxation in paramagnetic Ga(Mn)As
  quantum wells'', Phys. Rev. B {\bf 79}, 155201 (2009).
\bibitem{zhang1}P. Zhang, J. Zhou, and M. W. Wu, ``Multivalley spin
  relaxation in the presence of high in-plane electric fields in
  $n$-type GaAs quantum wells'', Phys. Rev. B
{\bf 77}, 235323 (2008).
\bibitem{jiang2}J. H. Jiang, M. W. Wu, and Y. Zhou, ``Kinetics of spin
  coherence of electrons in $n$-type InAs quantum wells under intense
  terahertz laser fields'', Phys. Rev. B
{\bf 78}, 125309 (2008).
\bibitem{zhang2}P. Zhang and M. W. Wu, ``Effect of nonequilibrium
  phonons on hot-electron spin relaxation in n-type GaAs quantum
  wells'', Europhys. Lett. {\bf 92}, 47009 (2010).
\bibitem{lu}C. L\"{u}, J. L. Cheng, and M. W. Wu, ``Hole spin
  dephasing in $p$-type semiconductor quantum wells'', Phys. Rev. B {\bf 73},
  125314 (2006).
\bibitem{yzhou}Y. Zhou, J. H. Jiang, and M. W. Wu, ``Electron spin
  relaxation in $p$-type GaAs quantum wells'', New J. Phys.
{\bf 11}, 113039 (2009).
\bibitem{zhang3} P. Zhang and M. W. Wu, ``Hole spin relaxation in
  [001] strained asymmetric Si/SiGe and Ge/SiGe quantum wells'', Phys. Rev. B {\bf 80}, 155311 (2009).
\bibitem{lu1}C. L\"u, U. Z\"ulicke, and M. W. Wu, ``Hole spin
  relaxation in $p$-type GaAs quantum wires investigated by
  numerically solving fully microscopic kinetic spin Bloch
  equations'', Phys. Rev. B
{\bf 78}, 165321 (2008).
\bibitem{lu2}C. L\"u, H. C. Schneider, and M. W. Wu, ``Electron spin
  relaxation in $n$-type InAs quantum wires'', J. Appl. Phys. {\bf 106}, 073703 (2009).
\bibitem{jiang3}J. H. Jiang, Y. Y. Wang, and M. W. Wu, ``Reexamination
  of spin decoherence in semiconductor quantum dots from the
  equation-of-motion approach'', Phys. Rev. B {\bf 77}, 035323 (2008).
\bibitem{jiang4}J. H. Jiang and M. W. Wu, ``Electron-spin relaxation
  in bulk III-V semiconductors from a fully microscopic kinetic spin
  Bloch equation approach'', Phys. Rev. B {\bf 79}, 125206 (2009).
\bibitem{cheng}J. L. Cheng and M. W. Wu, ``Spin diffusion/transport in
  $n$-type GaAs quantum wells'', J. Appl. Phys.
{\bf 101}, 073702 (2007).
\bibitem{zhang4}P. Zhang and M. W. Wu, ``Spin diffusion in Si/SiGe
  quantum wells: Spin relaxation in the absence of D'yakonov-Perel'
  relaxation mechanism'', Phys. Rev. B {\bf 79}, 075303 (2009).
\bibitem{flat}M. E. Flatt\'e, J. M. Bayers, and W. H. Lau, in {\it Spin
dynamics in semiconductors}, (Springer, Berlin, 2002).

\bibitem{glazov}M. M. Glazov and E. L. Ivchenko, ``Precession spin
  relaxation mechanism caused by frequent electron-electron
  collisions'', JETP Lett. {\bf 75},
403 (2002).
\bibitem{brand}M. A. Brand, A. Malinowski, O. Z. Karimov, P. A. Mrsden,
R. T. Harley, A. J. Shields, I. Farrer, D. A. Ritchie, and
M. Y. Simmons, ``Precession and motional slowing of spin evolution in
a high mobility two-dimensional electron gas'', Phys. Rev. Lett.
 {\bf 89}, 236601 (2002); W. J. H. Leyland, R. T. Harley, M. Henini, A. J.
Shields, I. Farrer, and D. A. Ritchie, ``Energy-dependent
electron-electron scattering and spin dynamics in a two-dimensional
electron gas'', Phys. Rev. B {\bf 77}, 205321 (2008).
\bibitem{ruan}X. Z. Ruan, H. H. Luo, Y. Ji, Z. Y. Xu, and V. Umansky,
  ``Effect of electron-electron scattering on spin dephasing in a
  high-mobility low-density two-dimensional electron gas'',
Phys. Rev. B {\bf 77}, 193307 (2008).
\bibitem{ohno}Y. Ohno, R. Terauchi, T. Adachi, F. Matsukura, and
H. Ohno, ``Electron spin relaxation beyond D'yakonov-Perel'
interaction in GaAs/AlGaAs quantum wells'', Physica E {\bf 6}, 817 (2000).
\bibitem{bronold}F. X. Bronold, A. Saxena, and D. L. Smith,
  ``Semiclassical kinetic theory of electron spin relaxation in
  semiconductors'', Phys. Rev. B {\bf 70}, 245210 (2004).
\bibitem{gio}G. F. Giulianni and G. Vignale, {\it Quantum Theory of the
Electron Liquid}, (Cambridge University Press, Cambrage, England, 2005).
\bibitem{ljiang} L. Jiang and M. W. Wu, ``Control of spin coherence in
  $n$-type GaAs quantum wells using strain'', Phys. Rev. B {\bf 72}, 033311 (2005).
\bibitem{holl}A. W. Holleitner, V. Sih, R. C. Myers, A. C. Gossard, D. D.
Awschalom, ``Dimensionally constrained D'yakonov-Perel' spin relaxation
in $n$-InGaAs channels: transition from 2D to 1D'', New J. Phys. {\bf 9}, 342 (2007).
\bibitem{zheng2}F. Zhang, H. Z. Zheng, Y. Ji, J. Liu, and G. R. Li,
  ``Spin dynamics in the second subband of a quasi--two-dimensional
  system studied in a single-barrier heterostructure by time-resolved
  Kerr rotation'', Europhys.
 Lett. {\bf 83}, 47007 (2008).
\bibitem{lu3}C. L\"u, J. L. Cheng, M. W. Wu, and I. C. da Cunha Lima,
  ``Spin relaxation time, spin dephasing time and ensemble spin
  dephasing time in $n$-type GaAs quantum wells'',
Phys. Lett. A {\bf 365}, 501 (2007).
\bibitem{stich1}D. Stich, J. Zhou, T. Korn, R. Schulz, D. Schuh,
 W. Wegscheider, M. W. Wu, and C. Sch\"uller, ``Effect of initial spin
 polarization on spin dephasing and the electron $g$ factor in a
 high-mobility two-dimensional electron system'', Phys. Rev. Lett. {\bf 98},
176401 (2007).
\bibitem{stich2}D. Stich, J. Zhou, T. Korn, R. Schulz, D. Schuh,
 W. Wegscheider, M. W. Wu, and C. Sch\"uller, ``Dependence of spin
 dephasing on initial spin polarization in a high-mobility
 two-dimensional electron system'', Phys. Rev. B {\bf 76},
205301 (2007).
\bibitem{zheng1}F. Zhang, H. Z. Zheng, Y. Ji, J. Liu, and G. R. Li,
  ``Electrical control of dynamic spin splitting induced by exchange
  interaction as revealed by time-resolved Kerr rotation in a
  degenerate spin-polarized electron gas'', Europhys.
 Lett. {\bf 83}, 47006 (2008).
\bibitem{korn}T. Korn, D. Stich, R. Schulz, D. Schuh, W. Wegscheider, and
C. Sch\"uller, ``Spin dynamics in high-mobility two-dimensional
electron system'', Adv. Solid State Phys. {\bf 48}, 143 (2009).
\bibitem{run}A. P. Dmitriev, V. Y. Kachorovskii, M. S. Shur,
  ``High-field transport in a dense two-dimensional electron gas in
  elementary semiconductors'', J. Appl. Phys.
{\bf 89}, 3793 (2001).
\bibitem{conwell}E. M. Conwell, {\it High Field Transport in Semiconductors},
(Pergamon, Oxford, 1972).

\bibitem{bap1}T. C. Damen, L. Vina, J. E. Cunningham, J. Shah, and
  L. J. Sham, ``Subpicosecond spin relaxation dynamics of excitons and
  free carriers in GaAs quantum wells'', Phys. Rev. Lett. {\bf 67}, 3432 (1991),
\bibitem{bap2}J. Wagner, H. Schneider, D. Richards, A. Fischer, and
  K. Ploog, ``Observation of extremely long electron-spin-relaxation
  times in $p$-type $\delta$-doped GaAs/Al$_{x}$Ga$_{1-x}$As double
  heterostructures'', Phys. Rev. B {\bf 47}, 4786 (1993).
\bibitem{bap3}H. Gotoh, H. Ando, T. Sogawa, H. Kamada, T. Kagawa, and
  H. Iwamura, ``Effect of electron--hole interaction on electron spin
  relaxation in GaAs/AlGaAs quantum wells at room temperature'', J. Appl. Phys. {\bf 87}, 3394 (2000).
\bibitem{bap4}T. F. Boggess, J. T. Olesberg, C. Yu, M. E. Flatt\'e,
and W. H. Lau, ``Room-temperature electron spin relaxation in bulk
InAs'', Appl. Phys. Lett. {\bf 77}, 1333 (2000).
\bibitem{bap5}S. Hallstein, J. D. Berger, M. Hilpert, H. C.
Schneider, W. W. R\"uhle, F. Jahnke, S. W. Koch, H. M. Gibbs, G. Khitrova, and
M. Oestreich,  ``Manifestation of coherent spin precession in
stimulated semiconductor emission dynamics'', Phys. Rev. B {\bf 56}, R7076 (1997).
\bibitem{bap6}P. Nemec, Y. Kerachian, H. M. van Driel, and
  A. L. Smirl, ``Spin-dependent electron many-body effects in GaAs'', Phys. Rev. B {\bf 72}, 245202 (2005).
\bibitem{bap7}H. C. Schneider, J.-P. W\"ustenberg, O. Andreyev, K. Hiebbner,
L. Guo, J. Lange, L. Schreiber, B. Beschoten, M. Bauer, and
M. Aeschlimann, ``Energy-resolved electron spin dynamics at surfaces
of $p$-doped GaAs'', Phys. Rev. B {\bf 73}, 081302 (2006).

\bibitem{yang}C. Yang, X. Cui, S.-Q. Shen, Z. Xu, and W. Ge, ``Spin
  relaxation in submonolayer and monolayer InAs structures grown in a
  GaAs matrix'', Phys. Rev. B {\bf 80}, 035313 (2009).

\bibitem{song}P. H. Song and K. W. Kim, ``Spin relaxation of
  conduction electrons in bulk III-V semiconductors'', Phys. Rev. B {\bf 66}, 035207 (2002).

%\bibitem{meier}F. Meier and B. P. Zakharchenya, {\it Optical Orientation}
%(North-Holland, Amsterdam, 1984).

\bibitem{rev}{\it Semiconductor Spintronics and Quantum Computation}, edited
by D. D. Awschalom, D. Loss, and N. Samarth (Springer-Verlag,
Berlin, 2002);{\it Spin Physics
in Semiconductors}, edited by M. I. D'yakonov (Springer,
Berlin, 2008), and references therein.


\bibitem{dzhi}R. I. Dzhioev, K. V. Kavokin, V. L. Korenev, M. V. Lazarev, B.
Y. Meltser, M. N. Stepanova, B. P. Zakharchenya, D. Gammon,
and D. S. Katzer, ``Low-temperature spin relaxation in $n$-type
GaAs'', Phys. Rev. B {\bf 66}, 245204 (2002).
\bibitem{kruss}M. Krau\ss, R. Bratschitsch, Z. Chen, S. T. Cundiff, and H. C.
Schneider, ``Ultrafast spin dynamics in optically excited bulk GaAs at
low temperatures'', Phys. Rev. B {\bf 81}, 035213 (2010).
\bibitem{shen}K. Shen, ``A peak in density dependence of electron spin
  relaxation time in $n$-type bulk GaAs in the metallic regime'', Chin. Phys. Lett. {\bf 26}, 067201 (2009).
\bibitem{app1}I. Appelbaum, B. Huang, and D. J. Monsma, ``Electronic
  measurement and control of spin transport in silicon'', Nature {\bf
447}, 295 (2007).
\bibitem{app2}B. Huang, L. Zhao, D. J. Monsma, and I. Appelbaum,
  ``35\% magnetocurrent with spin transport through Si'', Appl. Phys. Lett. {\bf 91}, 052501 (2007).
\bibitem{dress}G. Dresselhaus, ``Spin-orbit coupling effects in zinc blende structures'', Phys. Rev. {\bf 100}, 580 (1955).
\bibitem{ljiang1}L. Jiang, M. Q. Weng, M. W. Wu, and J. L. Cheng,
  ``Diffusion and transport of spin pulses in an $n$-type
  semiconductor quantum well'', J. Appl. Phys. {\bf 98}, 113702 (2005).
\bibitem{crooker}S. A. Crooker, D. L. Smith, ``Imaging spin flows in
  semiconductors subject to electric, magnetic, and strain fields'', Phys. Rev. Lett.
{\bf 94}, 236601 (2005).
\bibitem{beck}M. Beck, C. Metzner, S. Malzer, and G. H. D\"ohler,
  ``Spin lifetimes and strain-controlled spin precession of drifting
  electrons in GaAs'', Europhys. Lett. {\bf 75}, 597 (2006).

\bibitem{cheng1}J. L. Cheng, M. W. Wu, and I. C. da Cunha Lima,
  ``Anisotropic spin transport in GaAs quantum wells in the presence
  of competing Dresselhaus and Rashba spin-orbit coupling'', Phys. Rev. B {\bf 75}, 205328 (2007).

%\bibitem{rashba}Y. A. Bychkov and E. Rashba, ``Properties of a 2D
%  electron gas with lifted spectral degeneracy'', Pis???ma Zh. Eksp. Teor.
%Fiz. {\bf 39}, 66 (1984) [JETP Lett. {\bf 39}, 78 (1984)].


\bibitem{golub}N. S. Averkiev and L. E. Golub, ``Giant spin relaxation
  anisotropy in zinc-blende heterostructures'', Phys. Rev. B {\bf 60},
15582 (1999).
\bibitem{golub1}N. S. Averkiev, L. E. Golub, and M. Willander, ``Spin
  relaxation anisotropy in two-dimensional semiconductor systems'',
J. Phys.: Condens. Matt. {\bf 14}, R271 (2002).
\bibitem{loss}J. Schliemann, J.C. Egues, and D. Loss, ``Nonballistic
  spin-field-effect transistor'', Phys. Rev. Lett. {\bf 90}, 146801 (2003).
\bibitem{winkler} R. Winkler, ``Spin orientation and spin precession
  in inversion-asymmetric quasi-two-dimensional electron systems'', Phys. Rev. B {\bf 69}, 045317 (2004).
\bibitem{cheng2}J. L. Cheng and M. W. Wu, ``Spin relaxation under
  identical Dresselhaus and Rashba coupling strengths in GaAs quantum
  wells'', J. Appl. Phys. {\bf 99}, 083704
(2006).
\bibitem{footnote}$[\bar{1}10]$ or $[110]$ depends on the relative
  signs of the Dresselhaus and Rashba coupling strengths.

%\bibitem{helix1}B. A. Bernevig, J. Orenstein, and S.-C. Zhang, ``Exact
%  SU(2) symmetry and persistent spin helix in a spin-orbit coupled
%%  system'', Phys. Rev. Lett. {\bf 97},  236601 (2006).

%\bibitem{helix2}J. D. Koralek, C. P. Weber, J. Orenstein, B. A. Bernevig,
%S.-C. Zhang, S. Mack, and D. D. Awschalom, ``Emergence of the
%persistent spin helix in semiconductor quantum wells'', Nature {\bf 458}, 610 (2009).

\bibitem{shen1}K. Shen and M. W. Wu, ``Infinite spin diffusion length
  of any spin polarization along direction perpendicular to effective
  magnetic field from Dresselhaus and Rashba spin-orbit couplings with
  identical strengths in (001) GaAs quantum wells'', J. Supercond. Nov. Magn.
{\bf 22}, 715 (2009).

\bibitem{grat1} C. P. Weber, N. Gedik, J. E. Moore, J. Orenstein, J. Stephens,
and D. D. Awschalom, ``Observation of spin Coulomb drag in a
two-dimensional electron gas'', Nature {\bf 437}, 1330 (2005).
\bibitem{grat2}A. R. Cameron, P. Rickel, and A. Miller, ``Spin
  gratings and the measurement of electron drift mobility in multiple
  quantum well semiconductors'', Phys. Rev. Lett. {\bf 76},
4793 (1996).
\bibitem{grat3} C. P. Weber, J. Orenstein, B. A. Bernevig, S.-C. Zhang, J.
Stephens, and D. D. Awschalom, ``Nondiffusive spin dynamics in a two-dimensional electron gas'', Phys. Rev. Lett. {\bf 98}, 076604 (2007).

\bibitem{weng7}M. Q. Weng, M. W. Wu, and H. L. Cui, ``Spin relaxation
  in $n$-type GaAs quantum wells with transient spin grating'', J. Appl. Phys. {\bf
103}, 063714 (2008).

\end{thebibliography}

\end{document}